\documentclass[acmtist]{acmsmall} 

\usepackage{perpage}
\MakePerPage{footnote}
\usepackage{wrapfig}
\usepackage{balance}  
\usepackage{algorithm,algorithmic}
\usepackage{graphicx}
\usepackage{multirow}
\usepackage{array}
\usepackage{ctable}
\usepackage{filecontents}
\usepackage{epstopdf}
\usepackage[size=small,format=plain,labelfont=up,textfont=up]{caption}
\usepackage{mathtools}

\usepackage{subfigure}

\usepackage{comment}
\usepackage{footnote}
\usepackage{url}
\usepackage[
			colorlinks = cyan,
          	linkcolor = cyan,
            urlcolor  = cyan,
            citecolor = cyan,
            anchorcolor = cyan]{hyperref}
\makesavenoteenv{table}

\newcommand{\tn}{\tabularnewline}

\renewcommand{\footnotesize}{\scriptsize}

\long\def\MCMC {\textit{MTC}}

\begin{document}

\markboth{L. Tran et al.}{A Real-Time Framework for Task Assignment in Hyperlocal Spatial Crowdsourcing}

\title{A Real-Time Framework for Task Assignment in Hyperlocal \\ Spatial Crowdsourcing}
\author{Luan Tran\footnotemark[1]
\affil{University of Southern California}
Hien To\footnote{The authors wish it to be known that, in their opinion, the first two authors should be regarded as joint First Authors.}
\affil{University of Southern California}
Liyue Fan
\affil{University at Albany, SUNY}
Cyrus Shahabi
\affil{University of Southern California}
}

\begin{abstract}
\emph{Spatial Crowdsourcing (SC)} is a novel platform that engages individuals in the act of collecting various types of spatial data. This method of data collection can significantly reduce cost and turnover time, and is particularly useful in urban environmental sensing, where traditional means fail to provide fine-grained field data. In this study, we introduce hyperlocal spatial crowdsourcing, where all workers who are located within the spatiotemporal vicinity of a task are eligible to perform the task, e.g., reporting the precipitation level at their area and time. In this setting, there is often a $budget$ constraint, either for every time period or for the entire campaign, on the number of workers to activate to perform tasks. The challenge is thus to maximize the number of assigned tasks under the budget constraint, despite the dynamic arrivals of workers and tasks.   We introduce a taxonomy of several problem variants, such as \textit{budget-per-time-period} vs. \textit{budget-per-campaign} and \textit{binary-utility} vs. \textit{distance-based-utility}.   We study the hardness of the task assignment problem in the \textit{offline} setting and propose \textit{online} heuristics which exploits the spatial and temporal knowledge acquired over time.  Our experiments are conducted with spatial crowdsourcing workloads generated by the SCAWG tool and extensive results show the effectiveness and efficiency of our proposed solutions.

\end{abstract}

%
%



%
%


\acmformat{Luan Tran, Hien To, Liyue Fan, Cyrus Shahabi, 2017. A Real-Time Framework for Task Assignment in Hyperlocal Spatial Crowdsourcing.}



\maketitle

\section{Introduction}
\label{sec:intro}

With the ubiquity of smart phones and the improvements of wireless network bandwidth, every person with a mobile phone can now act as a multimodal sensor collecting and sharing various types of high-fidelity spatiotemporal data instantaneously.  
In particular, crowdsourcing for weather information has become popular.  
With a few recent apps, such as mPING\footnote{http://mping.nssl.noaa.gov/} and WeatherSignal\footnote{http://weathersignal.com}, individual users can report weather conditions, air pollutions, noise levels, etc.  In fact, the authors  in~\cite{Dorminey2014} regards crowdsourcing as ``the future of weather forecasting".

Through our collaboration with the Center for Hydrometeorology and Remote Sensing (CHRS)\footnote{http://chrs.web.uci.edu/} at the University of California, Irvine, we have developed a mobile app, iRain\footnote{https://itunes.apple.com/us/app/irain-uci/id982858283}~\cite{iRain}, to perform \textit{spatial crowdsourcing} for precipitation information. Unlike other weather crowdsourcing apps, iRain allows CHRS researchers to \emph{request} rainfall information at specific locations and times where their global satellite precipitation estimation technologies\footnote{http://hydis.eng.uci.edu/gwadi/} fail to provide real-time, fine-grained data.  Individual iRain users around those locations can \emph{respond} to those requests by reporting rainfall observations, e.g., heavy/medium/light/none.

\begin{wrapfigure}{r}{0.55\textwidth}
	\begin{center}
		\includegraphics[width=0.55\textwidth]{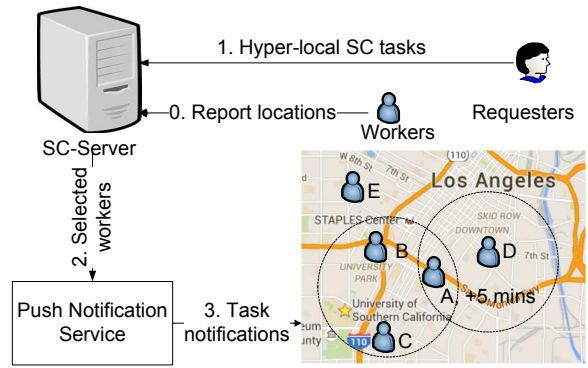}
	\end{center}
	\caption{Hyperlocal spatial crowdsourcing framework.}
	\label{fig:sc_framework}
\end{wrapfigure}

In general, spatial crowdsourcing (SC)~\cite{kazemi2012geocrowd} offers an effective data collection platform where data requesters can create spatial tasks dynamically and workers are assigned to tasks based on their locations. Figure~\ref{fig:sc_framework} depicts the architecture of iRain.   A requester issues a set of rainfall observation tasks to the SC-server (Step 1) where each task corresponds to a specific geographical extent, e.g., a circle.  The workers continuously update their locations to the SC-server when they become available for performing tasks (Step 0). Subsequently, the SC-server crowdsources the tasks among the workers in the task regions and sends the collected data back to the requester (Steps 2, 3). 

One major difference from existing SC paradigms~\cite{kazemi2012geocrowd,he2014toward,to2015server,xiao2015multi,guo2016activecrowd} is that workers in our paradigm do not need to travel to the exact task locations, e.g., to the centers of the circular regions, and are eligible to perform tasks as long as they are in close spatiotemporal vicinity of the tasks, e.g., enclosed in the circular regions\footnote{Tasks that require workers to physically travel to task locations, e.g., taking a picture of an event, are not considered in our problem setting.}.  We denote this new paradigm as \emph{Hyperlocal Spatial Crowdsourcing}.  The reason is two-fold.  Without requiring the workers travel physically, our paradigm lowers the threshold for worker participation and will potentially yield faster response.  Furthermore, the requested data, e.g., rainfall or temperature, exhibits spatiotemporal continuity in measurement.  Therefore, observations obtained at nearby locations, e.g., within a certain distance to the task location, and close to the requested time, are sufficient to fulfil the task.  For example, workers $B$ and $C$ in Figure~\ref{fig:sc_framework} are both eligible to report precipitation level at University of Southern California (USC), and worker $A$ who becomes available 5 minutes later is also qualified.  The acceptable ranges of space and time can be specified by data requesters, from which the SC-server can find the set of eligible workers for each task.


The SC-server operates to maximize fulfilled tasks for revenue.  Therefore it cannot assign tasks to an unlimited number of workers due to budget considerations.  In Hyperlocal SC, the \textit{budget} represents the payment to each selected worker upon task completion, or the communication cost for sending/receiving task notifications between the SC-server and each selected worker. 
Furthermore, it is not necessary to select many workers for overlapping tasks.  For example in Figure \ref{fig:sc_framework}, the observation of worker $A$ can be used for precipitation tasks at both USC and Los Angeles downtown (shown in two circles).


The goal of our study is to maximize the number of assigned tasks on the SC-server where only a given number of workers can be selected over a time period or during the entire campaign, i.e., under ``budget" constraints.   When tasks and workers are known \textit{a priori}, we can reduce the task assignment problem to the classic \emph{Maximum Coverage Problem} and its variants.  However, the main challenge with SC comes from the dynamism of the arriving tasks and workers, which renders an optimal solution infeasible in the online scenario. In Figure \ref{fig:sc_framework}, the SC-server is likely to activate worker $D$ and either worker $B$ or $C$ for the two tasks, respectively, without knowing that a more favorable worker $A$ is qualified for both tasks and will arrive in the near future. Previous heuristics in the literature~\cite{kazemi2012geocrowd,to2015server,Deng2016,guo2016activecrowd} do not consider the vicinity of tasks in space and time or the budget, thus cannot be applied to Hyperlocal SC. 

The contributions of this paper are as follows\footnote{This paper is an extension of a short paper appeared in~\cite{to2016real}.}
\noindent \textbf{1)} 
We provide a formal definition of Hyperlocal Spatial Crowdsourcing, where the goal is to maximize task coverage under budget constraints.
We introduce a taxonomy to classify several problem variants, e.g., given a budget constraint for each time period (${f\MCMC}$) vs. for the entire campaign (${d\MCMC}$). We show that both ${f\MCMC}$ and ${d\MCMC}$ variants are NP-hard in the offline setting.  
\noindent \textbf{2)} In the online setting, we propose several heuristics for real-time task assignment.  When a budget constraint is given for each time period (${f\MCMC}$), local heuristics i.e., \textit{Basic}, \textit{Temporal}, and \textit{Spatial},  are developed to select workers within each time period. When a budget is given for the entire campaign (${d\MCMC}$), we devise an adaptive strategy based on the contextual bandit to dynamically allocate the total budget to a number of time periods. 
\noindent \textbf{3)} When the utility of an assigned task is considered, we introduce two distance-based utility models to measure the assignment quality based on worker-task distance, which can be integrated with any previously developed heuristics.   To avoid overloading workers, we introduce a multi-objective variant in order to minimize the repetitive activation of the same worker.   Online solutions based on genetic algorithm and adaptive budget allocation are developed for ${f\MCMC}$ and ${d\MCMC}$ scenarios, respectively.    
\noindent \textbf{4)}
We conduct extensive experiments with workbench datasets generated from real-world location check-ins.  The empirical results confirm that our heuristics  are efficient and effective in assigning hyperlocal tasks in a real-time manner.

The remainder of this paper is organized as follows. Section \ref{sec:related} reviews the related work. Section \ref{sec:prelim} provides notations and a taxonomy for Hyperlocal SC problem.  We prove in Section \ref{sec:hardness} that offline task assignment in Hyperlocal SC with budget constraints is NP-hard.   In Section \ref{sec:online}, we study two problem variants in the online setting.  Section \ref{sec:overbooking} discusses the multi-objective optimization variant to mitigate worker overloading and Section~\ref{sec:task_value} describes the integration of distance-based task utility models. 
We report our experimental results in Section \ref{sec:exp}, provide discussion in Section~\ref{sec:discuss}, and conclude the paper in Section \ref{sec:con}.

\section{Related work}
\label{sec:related}
 
%

\textbf{Spatial Crowdsourcing (SC)} can be deemed as one of the main enablers of urban computing's applications such as monitoring traffic information and air pollution~\cite{zheng2014urban,ji2016urban}.
Only recently SC has gained popularity in both research community and industry, e.g., TaskRabbit, Gigwalk. A recent study \cite{to2015server} distinguishes SC from related fields, including generic crowdsourcing, participatory sensing, volunteered geographic information, and online matching. 
Research efforts in SC have focused on different aspects, such as \emph{task assignment} (e.g.,~\cite{kazemi2012geocrowd,Tong2016a,Liu2016,Cheng2016,Hu2016}), \emph{task scheduling} (e.g.,~\cite{Deng2016,Li2015,SalesFonteles2016}), \emph{quality control} and \emph{trust} (e.g., ~\cite{kazemi2013geotrucrowd,Cheng2015}), \emph{privacy} (e.g.,~\cite{to2014framework,to2017differentially,wanglocation}), \emph{incentive mechanism} (e.g.,~\cite{gao2015survey,Kandappu2016,zhang2017toward}).
The authors in \cite{kazemi2012geocrowd} proposed task assignment problem whose goal is to maximize the number of assigned tasks. Requesters may want to crowdsource a spatial \emph{complex} task that requires multiple workers at different locations to collectively perform several spatial sub-tasks~\cite{dang2013maximum,zhang2016capr,Gao2017}. 
In \cite{to2014framework}, the authors introduce the problem of protecting worker location privacy in SC. A framework is proposed to ensure differentially-private protection guarantees without significantly affecting the effectiveness and efficiency of the SC system.
Another study proposes a differentially private incentive mechanism in mobile crowd sensing system~\cite{jin2016enabling}. In such systems where workers bid tasks, worker's bid may reveal her interests, knowledge base. Thus, the proposed method preserves the privacy of each worker' bid against the other honest-but-curious workers. However, unlike our study, this work does not focus on challenges that are unique to spatial crowdsourcing such as spatial task allocation.
Meanwhile, the authors in~\cite{Pournajaf2014} propose a two-phase framework whose objective is to match a set of spatial tasks to a set of workers given the workers' cloaking regions such that task assignment is maximized while satisfying travel budget constraint of each worker.
The trust issues in SC have been studied in~\cite{kazemi2013geotrucrowd}, where one solution is having tasks performed redundantly by multiple workers. 
Recently in \cite{Cheng2016}, the reliability of task assignment is measured in term of both the confidence of task completion and the diversity quality of the tasks. 
However, the trust and reliability issues of workers are beyond the scope of our work; if there are multiple reports for one task, the SC-server will simply send all available reports to the task requester. 
It is worth noting that we assume the workers would respond to their assigned tasks, which is a common assumption in the server-assigned mode of spatial crowdsourcing~\cite{kazemi2012geocrowd}. There have been studies that relax this assumption by associating to each worker a probability to perform an assigned task (e.g.,~\cite{Cheng2016,to2014framework}). However, this consideration is not the focus of our paper. In addition, the time required for the workers responding to an assigned task is negligible (e.g., a few seconds with the iRain application) when compared to the deadline of each task (e.g., one day).

\textbf{Online Spatial Task Assignment:} 
There have been extensive studies regarding task assignment in generic crowdsourcing (e.g.,~\cite{venanzi2013crowdsourcing,tran2013efficient}).
However, unlike our study that focuses on the spatiotemporal aspects of the task assignment, these studies focus on task assignment in general crowdsourcing rather than spatial crowdsourcing.
Recent studies that are closely related to ours include~\cite{ul2014multi} and~\cite{Tong2016a}. Both of them study the online spatial task assignment problem; however, they differ from our work in terms of the problem setting and objectives. First, in our problem the report of a worker can be used for multiple tasks as long as their geographical extents cover the worker's location. Thus, our focus is worker selection rather than matching of workers to tasks as in~\cite{ul2014multi,Tong2016a}. Second, the objectives in these studies are respectively to maximize the number of assigned tasks and to maximize the total utility score of the worker-task matches while our framework considers both kinds of utility, each is jointly combined with other real-world considerations, i.e., leveraging historical workload and minimizing worker overloading. In addition, our study maximizes the utility of assignment under budget constraints while others focus solely on maximizing the utility.

\textbf{Budgeted Spatial Task Assignment:} 
There have been recent studies on matching workers with tasks under budget constraints.
In~\cite{tran2013efficient}, the authors propose CrowdBudget --- an agent-based budget allocation algorithm that divides a given budget among different tasks in order to achieve low estimation error (of the estimated answers for a set of tasks). This study does not consider the challenges of spatial task assignment, where workers and tasks can come and go at any time and we are not aware of their locations until their arrival time. 
The study in~\cite{miao2016balancing} also differs from ours. In our problem, the workers do not need to travel and report sensed value at their current locations. In contrast, in~\cite{miao2016balancing}, the workers do need to travel to the task locations, which may take a long time in rush hour. Consequently, the workers may reject their assigned tasks. It is worth noting that these studies focus on worker/task matching problem; however, our aim is to select the best workers to maximize task coverage. Similar to~\cite{zhang2015event,tran2013efficient}, the budget in~\cite{miao2016balancing} refers to payment to workers, while in our study, we consider budget as the number of workers to select and focus on allocating a total budget across multiple time periods.

\textbf{Worker Selection:}
Several works studied the problem of selecting workers with budget constraints~\cite{song2014qoi,zhang2014crowdrecruiter}. However, those studies focus on offline participant selection problem while our focus is to propose online solutions.
Furthermore, the problem settings in those studies differ from ours in several aspects.
Sensing tasks in~\cite{song2014qoi} are represented by non-overlapping regions while tasks in our study can overlap spatially thus more challenging for optimization. The authors in~\cite{zhang2014crowdrecruiter} studied the problem of selecting a minimum number of workers to minimize the overall incentive payment while satisfying a probabilistic coverage requirement; however, in our problem, the number of workers to be selected is constrained by a predefined budget.
 Our work is also closely related to the problem of matching workers with tasks~\cite{he2014toward,xiao2015multi}.  Particularly, in~\cite{he2014toward}, the authors studied the problem of task allocation that maximizes the reward of the SC platform given a time constraint for each worker. Recently in~\cite{xiao2015multi}, a task assignment problem that minimizes the average makespan of all assigned tasks was proposed.  Unlike these studies, SC workers in our setting need not travel to task locations.   Furthermore, our aim is different from the aforementioned studies, which is to maximize task coverage.

\section{Preliminaries}
\label{sec:prelim}
\begin{figure*}[ht]
	\centering
	\subfigure[Time period 1]{\label{fig:time_instance1}\includegraphics[width=.33\textwidth]{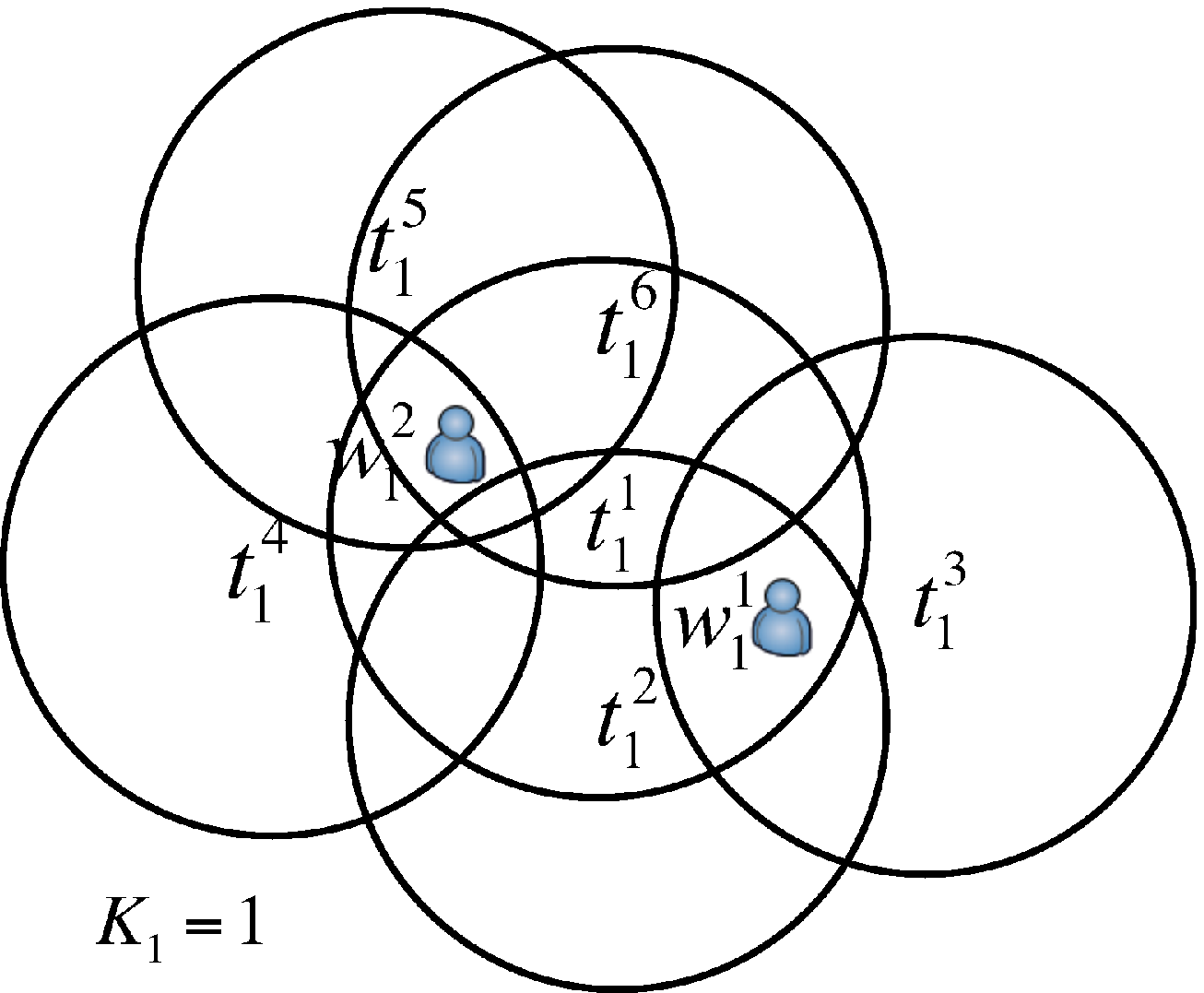}}
	\hspace{10pt}
	\subfigure[Time period 2]{\label{fig:time_instance2}\includegraphics[width=.33\textwidth]{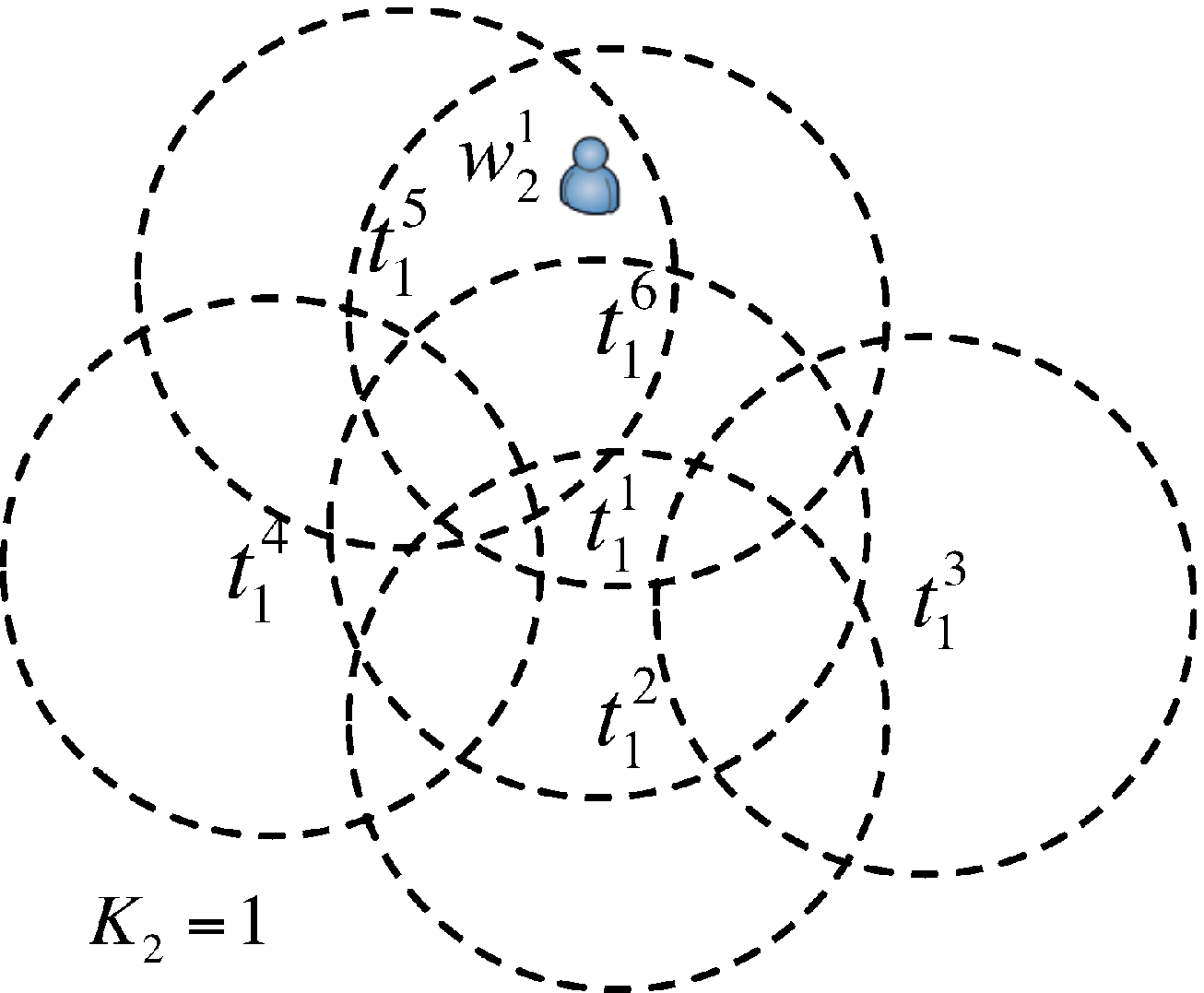}}
	\hspace{10pt}
	\subfigure[Bipartite graph]{\label{fig:bgraph}\includegraphics[width=.23\textwidth]{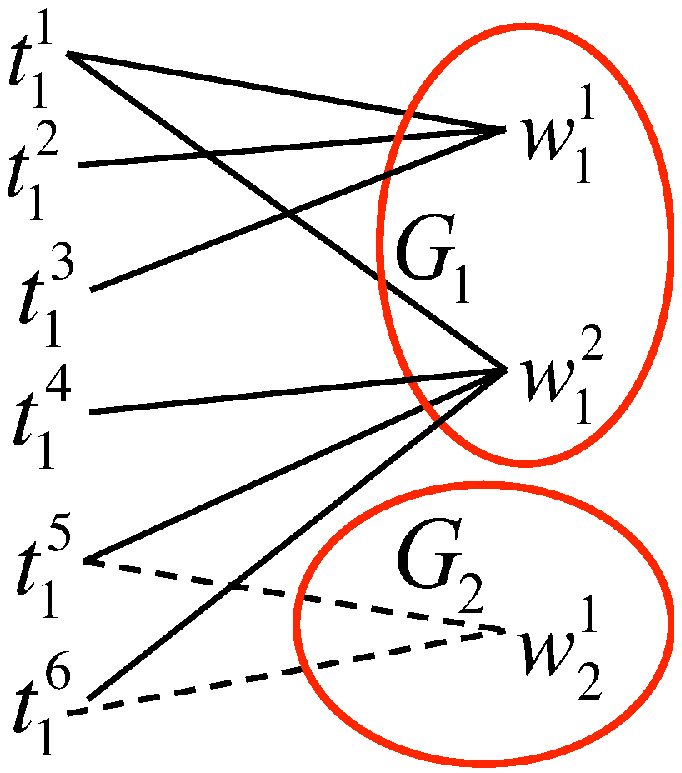}}
	\caption{Graphical example of worker-task coverage ($\delta=2$). Subscripts represent time periods while superscripts mean ids.}
	\label{fig:example}
\end{figure*}

We first introduce concepts and notations used in this paper.  A \textbf{task} is a query of certain hyperlocal information, e.g., precipitation level at a particular location and time. For simplicity, we assume that the result of a task is in the form of a numerical value, e.g., \emph{0=rain,1=snow,2=none}\footnote{Remote sensing techniques based on satellite images cannot differentiate between rain and snow.}.
Specifically, every task comes with a pre-defined region where any enclosed worker can report data for that task. In this paper, we define each task region as a circular space centered at the task location; however, task region can be extended to other shapes such as a polygon or to represent geography such as district, city, county, etc. 
Moreover, each task also specifies a valid \emph{time interval} during which users can provide data.
\begin{definition}[Task]
A task $t$ of form $<$${l,r,s,\delta}$$>$ is a query at location $l$, which can be answered by workers within a circular space centered at $l$ with radius $r$. The parameter $\delta$ indicates the duration of the query: it is requested at time $s$ and can be answered until time $s + \delta$.
\end{definition}
We refer to $s+\delta$ as the ``deadline'' of task $t$. A task expires if it has not been answered before its deadline.  Figure~\ref{fig:time_instance1} shows the regions of six tasks, $t_1^1,t_1^2,..., t_1^6$. All tasks expire at time period 2 (i.e., they can be deferred to time period 2), represented by the dashed circles in Figure~\ref{fig:time_instance2}. 

A \textbf{worker} can accept task assignments when he is \textit{online}.
\begin{definition}[Worker]
A worker $w$ of form $<$${id, l}$$>$, is a carrier of a mobile device who 
can accept spatial task assignments.
The worker can be uniquely identified by his $id$ and his location is at $l$.  
\end{definition}
Intuitively, a worker is eligible to perform a task if his location is enclosed in the task region. In Figure~\ref{fig:time_instance1}, $w_1^1$ is eligible to perform $t_1^1,t_1^2$ and $t_1^3$ while $w_1^2$ is qualified for $t_1^1,t_1^4,t_1^5$ and $t_1^6$.
Furthermore, a worker's report to one task can also be used for all other unexpired tasks whose task regions enclose the worker. As in Figure~\ref{fig:time_instance2}, $w_2^1$ is eligible to perform $t_1^5$ and $t_1^6$, which are deferred from time 1.




Let $W_i=\{w_i^1,w_i^2, ...\}$ denotes the set of available workers at time $s_i$  and $T_i=\{t_i^1,t_i^2, ...\}$ denotes the set of available tasks including tasks issued at time $s_i$ and previously issued un-expired tasks.  Below we define the notions of \textbf{worker-task coverage} and \textbf{coverage instance sets}.

\begin{definition}[Worker-Task Coverage]
Given $w_i^j \in W_i$, let $C(w_i^j) \subset T_i$ denotes the task coverage set of $w_i^j$, such that for every $t_i^k \in C(w^j)$,
\begin{align}
s_i < t_i^k.(s+\delta)  \label{eq:temporal} \\
||w_i^j.l - t_i^k.l ||_2 \le t_i^k.r \label{eq:spatial}
\end{align}
\end{definition}
We also say the worker $w_i^j$ covers the tasks $t_i^k \in C(w_i^j)$. An example of a coverage in Figure~\ref{fig:time_instance1} is $C(w_1^1) = \{t_1^1,t_1^2,t_1^3\}$.  

\begin{definition}[Coverage Instance Set]
At time $s_i$, the coverage instance set, denoted by $I_i$ is the set of worker-task coverage of form $<$$w_i^j,C(w_i^j)$$>$ for all workers $w_i^j \in W_i$.
\end{definition}
\begin{table}[h]
\begin{center}
\begin{tabular}{| l | p{6cm} |}
\hline
Time & Coverage Instance Sets \\
\hline
1 & $\{(w_1^1,$$<$$t_1^1,t_1^2,t_1^3$$>$$),(w_1^2,$$<$$t_1^1,t_1^4,t_1^5,t_1^6$$>$$)\}$ \\
\hline
2 & $\{(w_2^1,$$<$$t_1^5,t_1^6$$>$$)\}$ \\
\hline                                                                              
\end{tabular}
\caption {The coverage instance set of the example in Figure~\ref{fig:example}.}
\label{tab:coverage_set}
\vspace{-10pt}
\end{center}
\end{table}
The coverage instance sets for the example in Figure \ref{fig:example} are illustrated in Table~\ref{tab:coverage_set}.  For simplicity, we now assume the utility of a specific task assignment is \textit{binary} within the task region and before the deadline. That is, 
assignment to any worker within a task region before the deadline has utility 1 (1 successful assignment), and 0 otherwise.
As a result, task $t_1^5$ and $t_1^6$ being answered by worker $w_1^2$ at time $1$ is equivalent to it being answered by $w_2^1$ at time $2$.




Again, the goal of our study is to maximize task assignment given a budget, despite the dynamic arrivals of tasks and workers.
Now, we formally define the notion of a budget.
\begin{definition}[Budget]
Budget $K$ is the maximum number of workers to select in a coverage instance set.
\end{definition}
In practice, budget $K$ can capture the \emph{communication cost} the SC-server incurs to push notifications to selected workers (Step 3 in Figure \ref{fig:sc_framework}), or the \emph{rewards} paid to the workers. 

With  these,  we  formally  define  the hyperlocal crowdsourcing problem with budget constraint as follows.

\begin{definition}[Problem]
Given a set of workers $W$, a set of available tasks $T$, a budget constraint $K$, and a utility function $U(\cdot) \in \mathbb{R}$, find a subset of workers $W'$ of $W$ within the budget constraint, such that the total utility of the covered tasks is maximized.
\end{definition}

\subsection{Problem Taxonomy}

\subsubsection{Budget-per-time-period vs. Budget-per-campaign}
In certain scenarios, the task requester may specify a budget constraint, i.e., the maximum number of workers to select, for \textbf{each time period} in a campaign, e.g., a day or a week.   Given a set of time periods $\phi = \{s_1,s_2, ...,s_Q\}$, a budget constraint $K_i$ is specified for each $s_i$.  The challenge is to decide which workers to select within each time period.    On the other hand, the task requester could specify a budget constraint for the \textbf{entire campaign}.    Given a set of time periods $\phi = \{s_1,s_2, ...,s_Q\}$ and assuming $L_i$ workers are selected for $s_i$, a budget constraint $K$ is specified for the sum of $L_i$'s.   The new challenge of this problem variant is to allocate the total budget $K$ wisely over $Q$ time periods.  

The \textbf{choice} of the constraint model depends on the financial flexibility of the task requester.  Furthermore, the budget-per-time-period model is a special case of the budget-per-campaign model.  As a result, the utility of budget-per-campaign solution is no worse than that of the budget-per-period solution for any problem instance.


\subsubsection{Binary-utility vs. Distance-based-utility} Considering the utility of assigned tasks, our problem can be classified into binary-utility and distance-based-utility variants. In the \textbf{binary-utility} model, a task can be assigned to any worker located within the task radius to achieve utility $1$.  Unassigned tasks will yield $0$ utility.  Therefore, the optimization objective is to maximize the total number of assigned tasks.     However, for some applications, a worker who is closer to the task location may be ``preferred" over other workers farther away. For example, in weather crowdsourcing applications, e.g., iRain~\cite{iRain}, a closer worker can report more accurate rainfall data.  The \textbf{distance-based-utility} model thus evaluates a task assignment to a specific worker with various distance functions.

\subsubsection{Single-objective vs. Multi-objective}
In the \textbf{single-objective} problem formulation, we aim to maximize the total utility of assigned tasks.   On the other hand, crowdsourcing applications may have more than one, sometimes conflicting, objectives, to ensure long-term prosperity.  For example, worker overloading can be a critical concern of the novel crowdsourcing platforms, in which only a few workers are frequently selected to optimize task assignments.  Therefore, a \textbf{multi-objective} formulation can introduce a second objective to minimize the worker overloading phenomenon.  The challenge is thus to find solutions considering the trade-off between the two objectives.

\subsubsection{Offline vs. Online}
Orthogonal to the dimensions above, our problem can be further classified into offline and online variants.  The \textbf{offline} variant selects workers with complete knowledge of task/worker arrivals during the entire campaign.  Although this is not practical, studying the offline variant allows us to eliminate the hardness arising from the randomness of the online problem and focus on the optimization in a deterministic setting.  In the \textbf{online} variant, assignments have to be made in real-time for the currently arriving tasks/workers without complete knowledge of future arrivals.  While it is more fitting for crowdsourcing applications, it is also intuitively more challenging --- it is uncertain in nature when and where future tasks and workers may appear. Thus, effective worker selection must optimize the objective(s) in the long run.

The majority of this paper will focus on the online, binary-utility, single-objective problem with both per-time-period and per-campaign budget constraints.  We will also show how to extend our solutions to the distance-based utility model as well as the multi-objective problem.


\section{Hardness of the Problem}
\label{sec:hardness}

In this section we study the problem complexity of task assignment with budget constraint in hyperlocal spatial crowdsourcing. We show that two offline variants, i.e., budget-per-time-period vs. budget-per-campaign, of the problem are NP-hard and propose online heuristics in the next section.

\subsection{Fixed Budget ${f\MCMC}$}



\newtheorem{problem}{Problem}
\begin{problem}[Fixed-budget Maximum Task Coverage] \label{prob:fMTC}
Given a set of time periods $\phi = \{s_1,s_2, ...,s_Q\}$ and a budget $K_i$ for each $s_i$, the fixed-budget maximum task coverage ({f\MCMC}) problem is to select a set of workers $L_i$ at every $s_i$, such that the total number of covered tasks $|\bigcup_{i=1}^{Q}\bigcup_{w_i^j\in L_i}C(w_i^j)|$ is maximized and $|L_i| \le K_i$.
\end{problem}


This optimization problem is challenging since each worker is eligible for a subset of tasks. The fact that a task can be deferred to future time periods further adds to the complexity of the problem. With the following theorem, we proof that \textit{f\MCMC} is NP-hard by a reduction from the \emph{maximum coverage with group budgets} constraints problem (MCG) \cite{chekuri2004maximum}. MCG is motivated by the maximum coverage problem (MCP)~\cite{feige1998threshold}. Consider a given $I_g$, we are given the subsets $S=\{S_1,S_2,...S_m\}$ of a ground set $X$ and the disjoint sets $\{G_1,G_2,..., G_l\}$. Each $G_i$, namely a \emph{group}, is a subset of $S=\{S_1,S_2,...S_m\}$. With MCG, we are given an integer $k$, and an integer bound $k_i$ for each group $G_i$. A solution to $I_g$ is a subset $H \subset S$ such that $|H| \le k$ and $|H \cap G_i| \le k_i$ for $1\le i \le l$. The objective is to find a solution such that the number of elements of $X$ covered by the sets in $H$ is maximized. MCP is the special case of MCG. Since MCP is known to be strongly NP-hard~\cite{feige1998threshold}, by restriction, MCG is also NP-hard.

\newtheorem{thm}{Theorem}
\newtheorem{def_thm1}[thm]{Theorem}
\begin{def_thm1}
{f\MCMC} is NP-hard.
\end{def_thm1}
\begin{proof}
	We prove the theorem by a reduction from MCG~\cite{chekuri2004maximum}. That is, given an instance of the MCG problem, denoted by $I_g$, there exists an instance of the {\MCMC} problem\footnote{In this section, MTC refers to fixed-budget MTC for short}, denoted by $I_t$, such that the solution to $I_t$ can be converted to the solution of $I_g$ in polynomial time. The reduction has two phases, \emph{transforming} all workers/tasks across the entire campaign to a bipartite graph, and \emph{mapping} from MCG to {\MCMC}.
	First, we layout the tasks and workers as two set of vertices in a bipartite graph in Figure \ref{fig:bgraph}. A worker $w_i^j$ can cover a task $t_i^k$ if both spatial and temporal constraints hold, i.e., Equations \ref{eq:spatial} and \ref{eq:temporal}, respectively.
	In Figure \ref{fig:bgraph}, $w_2^1$ can cover $t_1^4$ and $t_2^5$, which are deferred from $s_1$ to $s_2$, represented by the dashed line.
	
	Thereafter, {\MCMC} can be stated as follows. Selecting the maximum $K_i$ workers per group, each group represents a time period, such that the number of covered tasks
	is maximized (i.e., $|L_i| \le K_i$). To reduce $I_g$ to $I_t$, we show a mapping from $I_g$ components to $I_t$ components. For every element in the ground set $X$ in $I_g$, we create a task $t_i^j$ ($1\le j \le |X|$). Also, for every set in $S$, we create a worker $w_i^j$ with $C(w_i^j=S_j)$ ($1\le j \le m$). Consequently, to solve $I_t$, we need to find a subset $L_i\subset W_i$ workers of maximum size $K_i$ in each group whose coverage is maximized. Clearly, if an answer to $I_t$ is the set $L_i$ ($1\le i \le Q$), the answer to $I_g$ will be the set $H \subset S$ of maximum coverage such that $|H| \le k=\sum_{i=1}^Q K_i$ and $|H \cap G_i| \le k_i = K_i$ for $1\le i \le Q$.
	
	As the transformation is bounded by the polynomial time to construct the bipartite graph, this completes the proof.
\end{proof}

By a reduction from the MCG problem, we can now use any algorithm that computes MCG to solve the {\MCMC} problem. 
The greedy algorithm in~\cite{chekuri2004maximum} provides $0.5$-approximation for MCG.  For example, the greedy solution in Figure~\ref{fig:bgraph} is $\{w_1^1,w_2^1\}$.  However, the approximation ratio only holds in the offline scenario where the server knows $apriori$ the coverage instance set for every time period.

\subsection{Dynamic Budget ${d\MCMC}$}

\begin{problem}[Dynamic-budget Maximum Task Coverage]
The dynamic-budget maximum task coverage problem (${d\MCMC}$), is similar to ${f\MCMC}$, except the total budget $K$ is specified for the entire campaign, i.e., $\sum_{i=1}^{Q}|L_i| \le K$.
\end{problem}


In the offline setting where the server is clairvoyant about the future workers and tasks, we prove the ${d\MCMC}$ problem is NP-hard by reduction from the maximum coverage problem (MCP).
\newtheorem{def_thm2}[thm]{Theorem}
\begin{def_thm2}
${d\MCMC}$ is NP-hard.
\end{def_thm2}
\begin{proof}
	We prove the theorem by a reduction from MCP. That is, given an instance of the MCP problem, denoted by $I_m$, there exists an instance of the {\MCMC} problem\footnote{In this section, MTC refers to dynamic-budget MTC for short}, denoted by $I_t$, such that the solution to $I_t$ can be converted to the solution of $I_m$ in polynomial time. The reduction includes two steps, \emph{transforming} all workers/tasks across the entire campaign to a bipartite graph, and \emph{mapping} from MCP to {\MCMC}. 
	The first step is similar to that of Theorem 1, in which the workers and tasks from the entire campaign are transformed into a bipartite graph as illustrated in Figure~\ref{fig:bgraph}. The \emph{mapping} step can be considered as a special case of the proof in Theorem 1, in which there exists only one group of all budget.
	As the transformation is bounded by the polynomial time to construct the bipartite graph and MCP is strongly NP-hard, this completes the proof.
\end{proof}

The results of these solutions to the offline scenarios will be used as the upper bounds of the results to the online solutions to be discussed in Section~\ref{sec:online}.
\section{Online Task Assignment}
\label{sec:online}
In this section we focus on \textit{online} variants: online \textit{f\MCMC} when a budget constraint is given for each time period, and online \textit{d\MCMC}, when the budget constraint is given for the entire campaign. We will introduce heuristics for each variant as follows.

\subsection{Fixed Budget ${f\MCMC}$}
\label{sec:fixed}

In the online scenario where workers and tasks arrive dynamically, it becomes more challenging to achieve the global optimal solution for Problem~\ref{prob:fMTC}.   Since the server does not have prior knowledge about future workers and tasks, it tries to optimize task assignment locally at every time period.  However, the optimization within every time period, similar to the \emph{maximum coverage problem} (MCP), is also NP-hard.  A greedy algorithm~\cite{feige1998threshold} was proposed to achieve an approximation ratio of $0.63$, by choosing a set which contains the largest number of \emph{uncovered} elements at each stage.  This study shows that the greedy algorithm is the best-possible polynomial time approximation algorithm for MCP.  Below we propose several greedy heuristics to solve the online ${f\MCMC}$ problem, namely \emph{Basic}, \emph{Spatial} and \emph{Temporal}.

%
%

\subsubsection{Basic Heuristic}

The \emph{Basic} heuristic solves the online ${f\MCMC}$ problem by using the greedy algorithm~\cite{hochbaum1996approximating} for every time period. At each stage, \emph{Basic} selects the worker that covers the maximum number of \emph{uncovered} tasks, depicted in Line~\ref{line:select_w} of Algorithm \ref{alg:basic}. For instance, in Figure \ref{fig:time_instance1}, $w_1^2$ is selected at the first stage. At the beginning of each time period, Line \ref{line:expired_task} removes expired tasks from the previous time period.
Line \ref{line:add_task} adds unassigned, unexpired tasks to current task set.
Line \ref{line:result} outputs the covered tasks $C_i$ per time period which will be used as the main performance metric in Section~\ref{sec:exp}. The algorithm terminates when either running out of budget or all the tasks are covered (Line \ref{line:termination}).

  \begin{wrapfigure}{R}{0.65\textwidth}
    \begin{minipage}{0.65\textwidth}
\begin{algorithm} [H]
\caption{\sc Basic Algorithm}
\small
\begin{algorithmic}[1]
\STATE Input: worker set $W_i$, task set $T_i$, budgets $K_i$
\STATE Output: selected workers $L_i$ \label{line:output}
\STATE For each time period $s_i$ \label{line:for_instance}
\STATE \hspace{10pt} Remove expired tasks $U_{i-1}'\leftarrow U_{i-1}$ \label{line:expired_task}
\STATE \hspace{10pt} Update task set $T_i\leftarrow T_i\cup U_{i-1}'$ \label{line:add_task}
\STATE \hspace{10pt} Remove tasks that do not enclose any worker $T_i'\leftarrow T_i$ \label{line:dummy_task}
\STATE \hspace{10pt} Construct worker set $W_i$, each $w_i^j$ contains $C(w_i^j)$ \label{line:mcmc_input}
\STATE \hspace{10pt} Init $L_i = \{\}$, uncovered tasks $R = T_i'$ \label{line:init_params}
\STATE \hspace{10pt} While $|L_i| < K_i$ and $|R| > 0$\label{line:termination}
\STATE \hspace{20pt} Select $w_i^j \in W_i - L_i$ that maximize $|C(w_i^j)\cap R|$  \label{line:select_w}
\STATE \hspace{20pt} $R\leftarrow R-C(w_i^j)$; $L_i\leftarrow L_i \cup \{w_i^j\}$ \label{line:update_covered_tasks}
\STATE \hspace{10pt} $C_i\leftarrow \bigcup_{w_i^j \in L_i} C(w_i^j)$ \label{line:result}
\STATE \hspace{10pt} Keep uncovered tasks $U_i\leftarrow T_i'-C_i$ \label{line:uncovered_task}
\end{algorithmic}
\label{alg:basic}
\end{algorithm}
    \end{minipage}
  \end{wrapfigure}

\emph{Basic} can achieve fast task assignment by simply counting the number of tasks covered by each worker (Line~\ref{line:select_w}).  However, it treats all tasks equally without considering the spatial and temporal information of each task, i.e., location and deadline. For example, a task located in a ``worker-sparse'' area may not be assigned in the future due to lack of nearby workers and thus should be assigned with higher priority at the current iteration.  Similarly, tasks that are expiring soon should be assigned with higher priorities.  Consequently, the priority of a worker is high if he covers a larger number of high priority tasks.  
Below we introduce two assignment heuristics that explicitly model the task priority given its spatial and temporal characteristics.  

\subsubsection{Temporal Heuristic}

One approach to prioritizing tasks is by considering their temporal urgency. The intuition is that a task which is further away from its deadline is more likely to be covered in the future, and vice versa.  
As a result, near-deadline tasks should have higher priorities to be assigned than others.  Consequently, a worker who covers a large number of soon-to-expire tasks should be preferred for selection. Based on the above intuition, we model the priority of a worker $w_i^j$ based on the remaining time of each task he covers as follows.
\begin{equation}
priority(w_i^j)=\sum_{t_i^k \in C(w_i^j)\cap R}\frac{1}{t_i^k.(s+\delta)-i}
\label{eq_ctdp}
\end{equation}
The \emph{Temporal} heuristic adapts \emph{Basic} by selecting the worker with maximum \emph{priority} at each stage. For instance, given two workers $w_1^1$ and $w_1^2$ at time $s_1$, where $C(w_1^1) = \{t_1^1, t_1^2\}$ and $C(w_1^1) = \{t_1^3\}$.  Suppose both $t_1^1$ and $t_1^2$ expire in $5$ time periods and $t_1^3$ expires in $2$ time periods.  The \emph{Temporal} heuristic chooses $w_1^2$ over $w_1^1$ as their priorities are $0.5$ and $0.4$, respectively. 
To implement \emph{Temporal}, Line 10 in Algorithm~\ref{alg:basic} can be updated to select the worker with maximum priority defined as in Equation~\ref{eq_ctdp}. We will empirically evaluate all heuristics in Section~\ref{sec:exp}.

\subsubsection{Spatial Heuristic}
\label{sec:spatial_heuristic}
To maximize task assignment in the long term, we also consider the ``popularity'' of a task location as an indicator of whether the task can be assigned to future workers.   Accordingly, we can spend the budget for the current time period to assign those tasks which can be only covered by existing workers.    The ``popularity'' of a task region can be measured using Location Entropy \cite{cranshaw2010bridging}, which captures the diversity of visits to that region.
A region has a high entropy if many workers visit with equal probabilities.
In contrast, a region has a low entropy if there are only a few workers visiting. 
We define the \emph{entropy} of a given task as follows.

\begin{wrapfigure}{r}{0.3\textwidth}
	\begin{center}
		\includegraphics[width=0.3\textwidth]{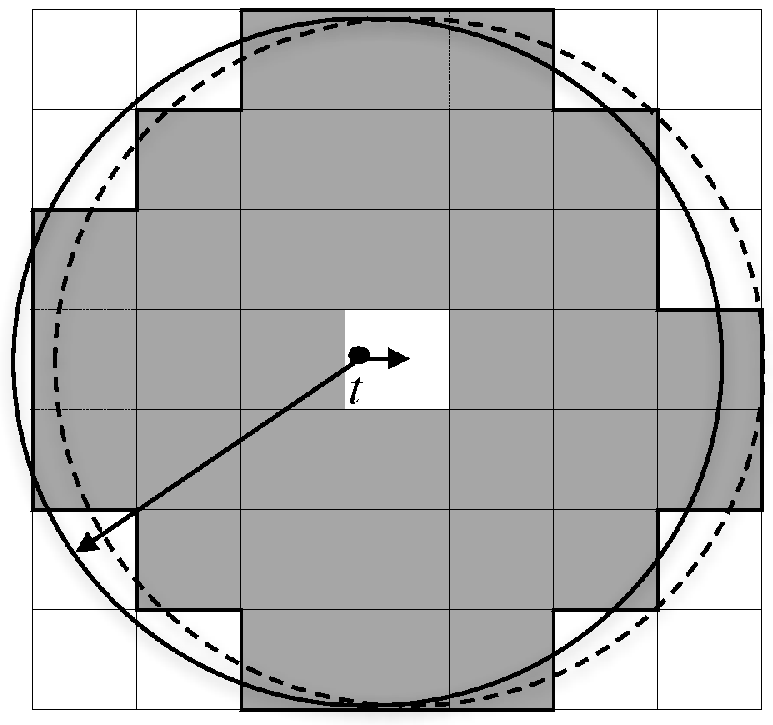}
	\end{center}
	\caption{Approximation of Task Entropy.}
	\label{fig:region_entropy}
\end{wrapfigure}

For task $t$, let $O_t$ be the set of visits to the task region $R(t.l,r)$. Let $W_t$ be the set of distinct workers that visited $R(t.l,r)$, and $O_{w,t}$ be the set of visits that worker $w$ made to $R(t.l,r)$. The probability that a random draw from $O_t$ belongs to $O_{w,t}$ is $P_t(w)=\frac{|O_{w,t}|}{|O_t|}$. The entropy of $t$ is computed as follows
\begin{equation}
\label{eq:region_entropy}
RE(t) = - \displaystyle\sum\limits_{w\in W_t}P_t(w)\times logP_t(w)
\end{equation}
For efficient evaluation, $RE(t)$ can be approximated by aggregating the entropies of 2D grid cells within the task region $R(t.l,r)$ and the cell entropies can be precomputed using historical data. Since any worker located inside $R(t.l,r)$ can perform task $t$,  $t$ is likely to be covered in the future as long as one grid cell inside $R(t.l,r)$ is ``popular'' among workers.
Figure~\ref{fig:region_entropy} illustrates the pre-computation of the entropy of task $t$. 
When a task arrives, we first identify the grid cell that encloses the task location, i.e., the white cell in the center, and slightly adjust the task region (solid circle) to be centered at the white cell (dashed circle).  We approximate the task entropy by the entropy of the dashed circle, which can be computed. This is because the dashed circle is solely determined by the white cell and radius $r$. To further speed up the precomputation of all possible combination of the cell and the radius, we approximate the dashed circle by a set of cells whose centers are within the circle.
With the entropy of every task covered by worker $w_i^j$, his priority can be calculated as follows
\begin{equation}
\label{eq:weight_re}
priority(w_i^j)=\sum_{t_i^k \in C(w_i^j)\cap R}\frac{1}{1 + RE(t_i^k)}
\end{equation}
Note that the constant $1$ is needed to avoid division by zero. Consequently, the \emph{Spatial} heuristic greedily selects the worker with maximum \emph{priority} at each stage. Line 10 in Algorithm \ref{alg:basic} can be modified to reflect the spatial priority of each worker.

\subsection{Dynamic Budget ${d\MCMC}$}
\label{sec:dynamic}
The second problem variant we study is more general, where a budget constraint is given for the entire campaign. This relaxation often results in higher task coverage. For example, in Figure \ref{fig:example}, if budget $1$ is given at every time period, we select $w_1^1$ and $w_2^1$ and obtain the coverage of 5. However, the dynamic-budget variant yields higher coverage of 6 by selecting $w_1^1$ and $w_1^2$ at time $1$.  Below we study the problem complexity in the offline scenario and propose adaptive budget allocation strategies for the online scenario.


The challenge of the online $d{\MCMC}$ problem is to allocate the overall budget $K$ over $Q$ time periods ($K\ge Q$) optimally, despite the dynamic arrivals of workers and tasks.  Below we introduce  several budget allocation strategies.  Once a budget is allocated to a particular time period, we can adopt previously proposed heuristics, i.e., \emph{Basic, Spatial, Temporal}, to select the best worker.  




The simplest strategy, namely \emph{Equal}, equally divides $K$ to $Q$ time periods; each time period has $K/Q$ budget and the last time period obtains the remainder.
However, \emph{Equal} may over-allocate budget to the time periods with small numbers of tasks. 
Another strategy is to allocate a budget to each time period proportional to the number of available tasks at that time period, i.e., $\frac{|T_i|}{|T|}K$, where $|T|$ is the total number of tasks. However, $|T|$ is not known \textit{a priori}. Furthermore,  we may still over-allocate budget to any time period with large $|T_i|$, if none of the tasks can be covered by any workers (or all the tasks can be covered by 1 worker).   We cannot allocate budget optimally without looking at the coverage instance set at each time period.

\subsubsection{Adaptive Budget Allocation}
\label{sec:adaptive_budget}

To maximize task assignment, we need to adaptively allocate the overall budget and consider the ``return" of selecting every worker, i.e., the worker priority, given the dynamic coverage instance set at every time period.  
We define the following two notions.  \textbf{Delta budget}, denoted as $\delta_K$, captures the current status of budget utilization, compared to a baseline budget strategy $\{K^{base}[t], t = 1, \dots, Q\}$, e.g., the \emph{Equal} strategy . Given a certain baseline $\{K^{base}[t]\}$, $\delta_K$ is the difference between the cumulative baseline budget and the actual budget spent up to time period $s_i$.   Formally, at any time period $s_i$,
\begin{align}
\delta_K = \sum_{t=1}^{i}(K^{base}[t])-K_{used}\label{eq:delta_budget}
\end{align}
A positive $\delta_K$ indicates budget is under-utilized, and vice versa.  Another notion is \textbf{delta gain}, denoted as $\delta_\lambda$, which represents the return of a worker currently being considered ($\lambda_l$) compared to the ones selected in the past ($\overline{\lambda_{l-1}}$). Formally,
\begin{align}
\delta_\lambda=\lambda_l-\overline{\lambda_{l-1}}
\label{eq:deltaGain}
\end{align}
where $\lambda_l$ is the gain of the current worker, calculated by any previously proposed local heuristic, i.e., as $|priority(w_i^j)|$.  $\overline{\lambda_{l-1}}$ is the average gain of previously added workers, i.e., $\overline{\lambda_{l-1}}=\frac{1}{l-1}\sum_{t=1}^{l-1}\lambda_t$.  A positive $\delta_\lambda$ indicates the current worker has higher priority than the historical average, and vice versa.  

Based on the contextual information $\delta_K$ and $\delta_\lambda$ at each stage of worker selection, we examine all available workers at the currently time period and decide whether to allocate budget $1$ to selecting any worker.  Intuitively, when both $\delta_K$ and $\delta_\lambda$ are positive, i.e., the budget is under-utilized and a worker has higher priority, the selection of the considered worker is favored.  When both are negative, it may not be worthwhile to spend the budget.   The other cases when one is positive and the other is negative are more complex, as we would like to spend budget on workers with higher priority but also need to save budget for future time periods in case better worker candidates arrive. 

\begin{figure}[!ht]\centering
	\includegraphics[width=0.8\textwidth]{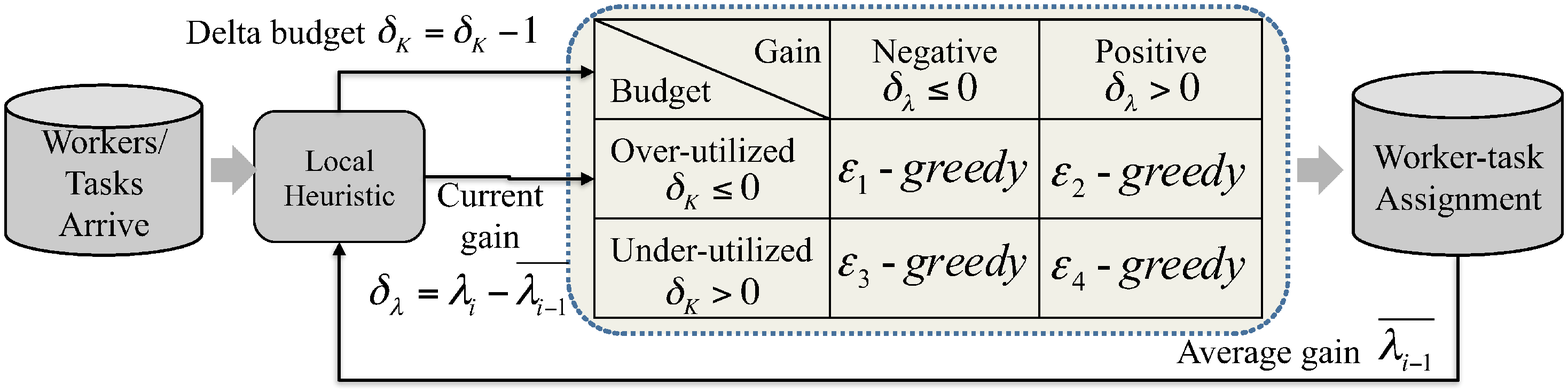}
	\caption{Adaptive budget allocation based on contextual $\epsilon$-greedy.}
	\label{fig:adaptive_framework}
\end{figure}

 
Our solution to the sequential decision problem is inspired by the well-known multi-armed bandit problem (MAB),
which has been widely studied and applied to decisions in clinical trials,
online news recommendation,
and portfolio design.
$\epsilon$-greedy, which achieves a trade-off between exploitation and exploration, proves to be often hard to beat by other MAB algorithms~\cite{vermorel2005multi}.  Hence, we propose an adaptive budget allocation strategy, based on \textit{contextual $\epsilon$-greedy} algorithm~\cite{li2010contextual}.
We illustrate our solution in Figure~\ref{fig:adaptive_framework}.  

At each stage of the local heuristic, a binary decision to make is whether to allocate budget $1$ to activate the current worker with the highest priority.   The contextual $\epsilon$-greedy algorithm allows us to specify an exploration-exploitation ratio, i.e., $\epsilon$, based on the worker's context, i.e., $\delta_K$ and $\delta_\lambda$.  As depicted in Figure~\ref{fig:adaptive_framework}, an $\epsilon_i$-greedy algorithm is used to determine whether to select the current worker based on his $\delta_K$ and $\delta_\lambda$.    For each case, a YES decision is made with $1-\epsilon_i$ probability and a NO decision with $\epsilon_i$ probability.  By default, we set  $\epsilon_1=1$ and $\epsilon_4=0$ to reflect NO and YES decisions, respectively, as discussed before.  When $\delta_K$ and $\delta_\lambda$ have different signs, the decision is not as straightforward as the other cases and thus we set $\epsilon_2=\epsilon_3=0.5$ to allow YES and NO decisions with equal probabilities. 
The pseudo code of our adaptive algorithm is depicted in Algorithm~\ref{alg:adapt}.

  \begin{wrapfigure}{R}{0.65\textwidth}
    \begin{minipage}{0.65\textwidth}
\begin{algorithm} [H]
	\caption{\sc Adaptive Budget Algorithm (Adapt)}
	\small
	\begin{algorithmic}[1]
		\STATE Input: $W_i$, $T_i$, total budgets $K$
		\STATE Output: selected workers $L_i$ \label{line:ad_output}
		\STATE Init $R = T_i$; used budget $K_{used} = 0$; average gain $\overline{\lambda_{i-1}} = 0$
		\STATE Budget allocation $K^{equal}[]$ with Equal strategy \label{line:ad_equal_budget}
		\STATE For each time period $s_i$ \label{line:ad_for_instance}
		\STATE \hspace{10pt} Perform Lines \ref{line:expired_task}-\ref{line:init_params} from Algorithm \ref{alg:basic}
		\STATE \hspace{10pt} Remained budget $K_i = K-K_{used}$ \label{line:ad_start_stats}
		\STATE \hspace{10pt} If $i=Q$, then $\delta_K = K_i$ \COMMENT{the last time period}
		\STATE \hspace{10pt} Otherwise, $\delta_K = (\sum_{t=1}^{i}K^{equal}[t]) -  K_{used}$ \label{line:ad_end_stats}
		\STATE \hspace{10pt} While $|L_i| < K_i$ and $R$ is not empty: \label{line:ad_termination}
		\STATE \hspace{20pt} Select $w_i$ in $W_i$ with highest $\lambda_i$  \label{line:ad_select_w}
		\STATE \hspace{20pt} Delta gain $\delta_\lambda=\lambda_i-\overline{\lambda_{i-1}}$
		\STATE \hspace{20pt} If $\delta_\lambda \le 0$ and $\delta_K \le 0$ and $rand(0,1) \le \epsilon_1$, then break
		\STATE \hspace{20pt} If $\delta_\lambda \le 0$ and $\delta_K > 0$ and $rand(0,1) \le \epsilon_2$, then break
		\STATE \hspace{20pt} If $\delta_\lambda > 0$ and $\delta_K > 0$ and $rand(0,1) \le \epsilon_3$, then break
		\STATE \hspace{20pt} If $\delta_\lambda > 0$ and $\delta_K \le 0$ and $rand(0,1) \le \epsilon_4$, then break
		\STATE \hspace{20pt} $\delta_K = \delta_K - 1$ \label{line:ad_update_budget_status}
		\STATE \hspace{20pt} Perform Line \ref{line:update_covered_tasks} from Algorithm \ref{alg:basic}
		\STATE \hspace{10px} $K_{used} = K_{used} + |L_i| $ \COMMENT{update the budget}
		\STATE \hspace{10px} $\overline{\lambda_{i}} = (\overline{\lambda_{i-1}}(Q-1) + \lambda_{i})/Q$
		\STATE \hspace{10pt} Perform Lines \ref{line:result},\ref{line:uncovered_task} from Algorithm \ref{alg:basic}
	\end{algorithmic}
	\label{alg:adapt}
\end{algorithm}
    \end{minipage}
  \end{wrapfigure}

\subsubsection{Historical Workload}

\begin{figure*}[ht]
	\centering
	\subfigure[Foursquare, 16x24 hours]{\label{fig:fo_cycle}\includegraphics[width=.37\textwidth]{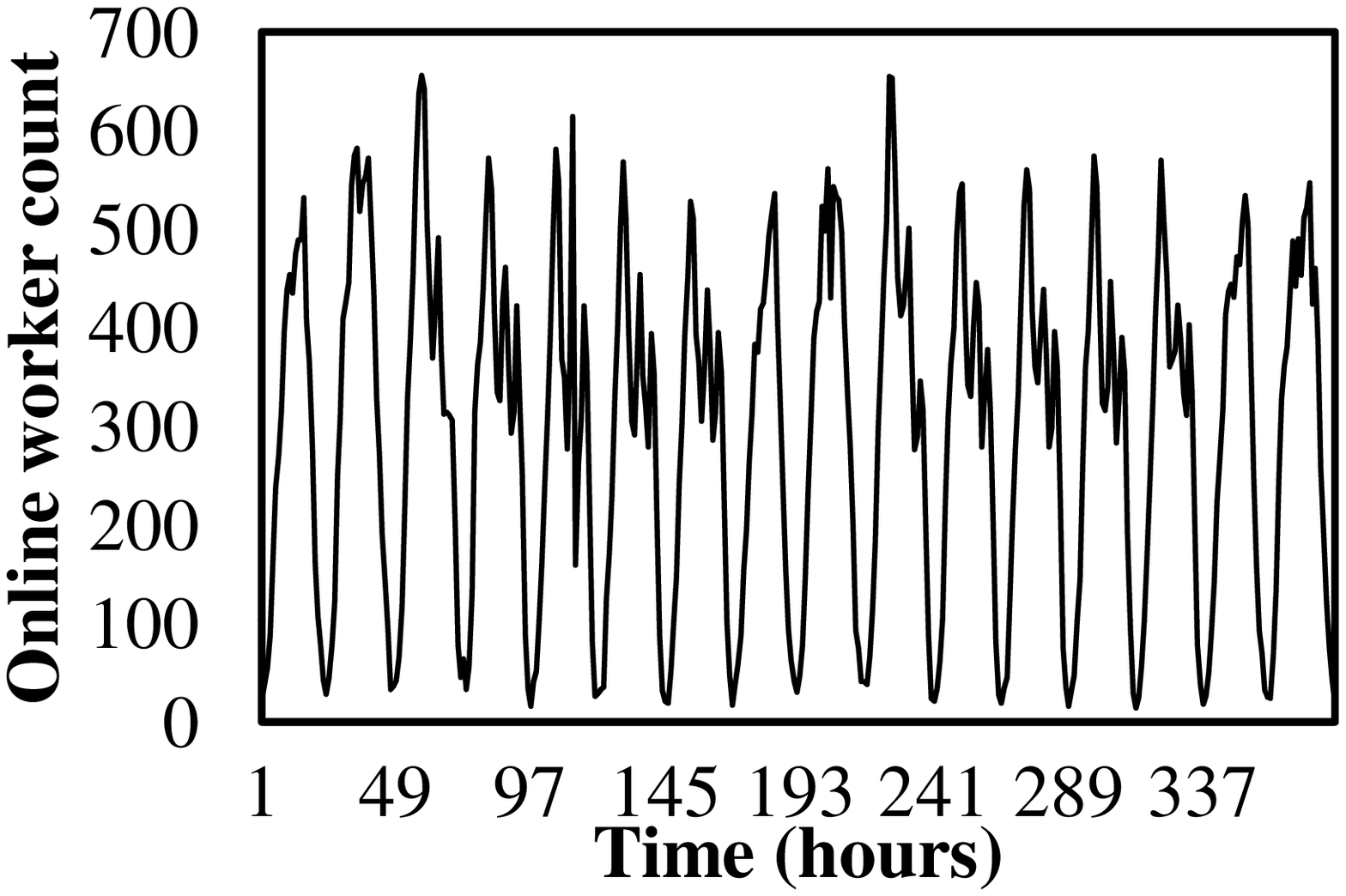}}
	\hspace{30pt}
	\subfigure[Gowalla, 32x7 days]{\label{fig:go_cycle}\includegraphics[width=.37\textwidth]{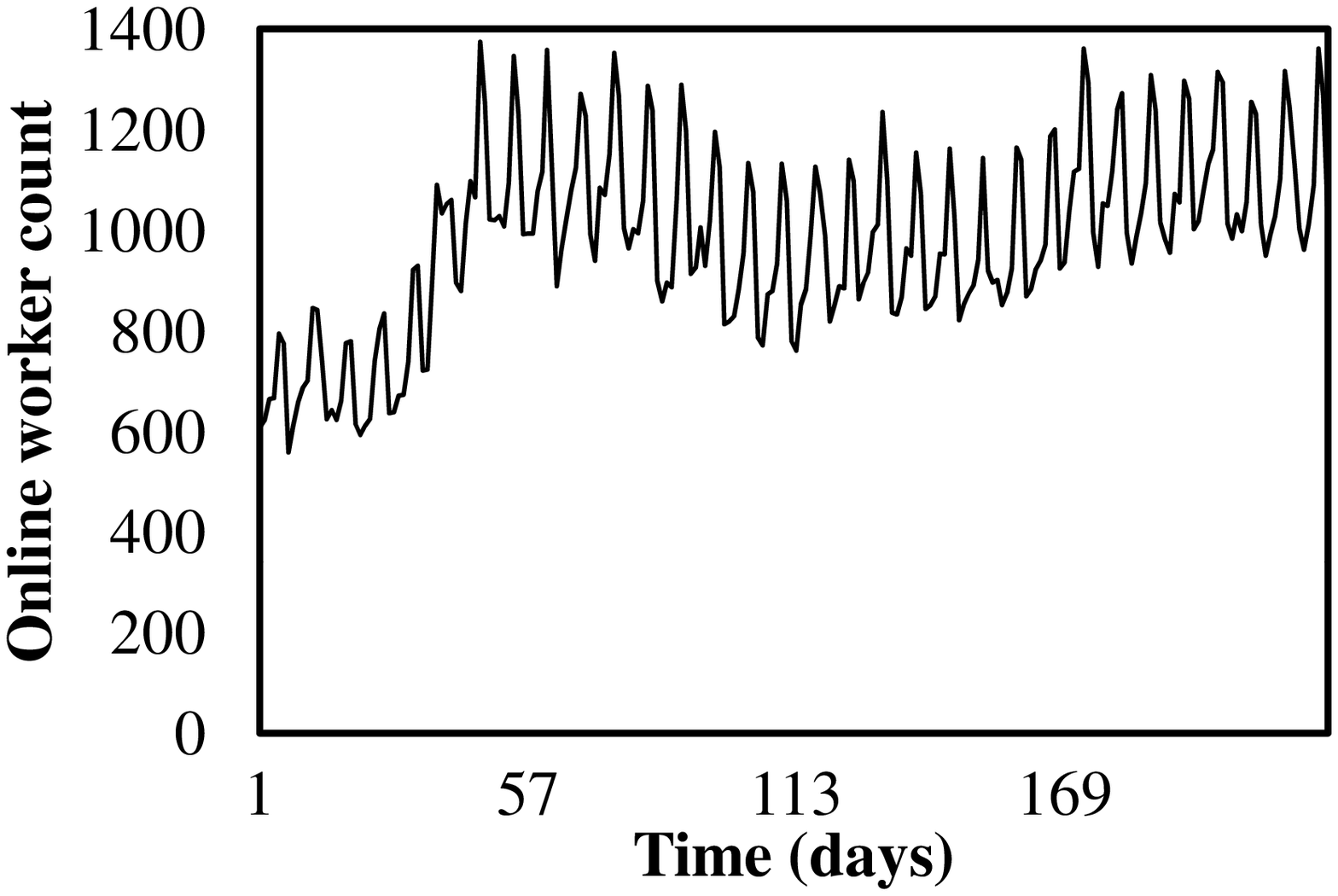}}
	\caption{Daily and weekly human activity patterns.}
	\label{fig:cycle_online_worker}
\end{figure*}

Previously our solution is simplified by considering $\{K^{equal}[t]\}$ as the baseline budget strategy. Since human activity exhibits temporal patterns, understanding those patterns may help to guide budget allocation. Therefore, we propose to compute a baseline budget strategy with historical data that captures the expected worker/task patterns.
The study in~\cite{musthag2013labor} shows the time-of-day usage patterns of workers in mobile crowdsourcing applications. The activity peaks are between 4 to 7 pm when workers leave their day jobs. Similar patterns are observed in Foursquare and Gowalla data sets in Figure~\ref{fig:cycle_online_worker}. Figure~\ref{fig:fo_cycle} shows the hourly count of check-ins present three peaks, i.e., during lunch and morning/afternoon commute.  In Figure~ \ref{fig:go_cycle}, we can observe peak check-in activities during weekends.

With historical worker and task information, we can leverage the optimal budget allocation strategy in the recent past and use it as the baseline strategy in Equation~\ref{eq:delta_budget}.  We propose to learn the budget allocation of previous time periods, namely \emph{workload}, using the greedy algorithm for the offline $d\MCMC$ problem.  To guide future budget allocation decisions, the previous workload $K^{prev}[]$ will be used as the baseline in Equation~\ref{eq:delta_budget}.  We will empirically evaluate our proposed solutions in the experiment section.  
\section{Worker Overload}
\label{sec:overbooking}
In this section, we present an enhancement to our solution in order to avoid repetitive activations of the same workers.  
The practical implication is that those workers who locate in popular areas can be repeatedly selected by our heuristics. Overloading workers may result in undesirable consequences, such as tasks being rejected and the workers either feel annoyed or stressed out to report.
Several recent studies~\cite{alfarrarjeh2015scalable,zhang2016capr,liu2016cost} also discuss the issue of over-assigning tasks to workers. These studies minimize worker overloading by balancing the workload of the workers. For example, the objective is to find an assignment that minimizes the variance of the workload among workers, i.e., maximize the so-called social fairness~\cite{liu2016cost}. Another work~\cite{alfarrarjeh2015scalable} also aims to assign a similar number of tasks to each worker. However, none of these studies considers task assignment and worker overloading as a multi-objective optimization problem.

Our idea is to minimize the phenomenon of overloading.
This requires to maintain the number of times each worker $w_{id}$ has been activated up to time $i$, $map$$<$$w_{id},count_i(w_{id})$$>$. The counter is defined as:
\setlength{\belowdisplayskip}{0pt} \setlength{\belowdisplayshortskip}{0pt}
\setlength{\abovedisplayskip}{0pt} \setlength{\abovedisplayshortskip}{0pt}
\begin{align}
count_i(w_{id}) = \sum_{k=1}^i \sum_{j=1}^{|W_k|} d(w_k^j) [w_k^j.id = {id}]
\end{align}

where $d(w_k^j)$ represents a decision to select the $j^{th}$ worker at time $k$: $d(w_k^j)=1$ if the worker is selected, otherwise $d(w_k^j)=0$.  The brackets enclose a condition that includes the term $d(w_k^j)$ to the sum \textit{iff} $w_k^j$ is identified by the same $id$.


We include minimization of worker overloading as another objective to coverage maximization. In the following, we formulate a multi-objective optimization (MOO) problem and propose solutions in both fixed-budget and dynamic-budget scenarios.

\subsection{Fixed Budget $f\MCMC$}
\label{sec:ol_fixed}
In the  fixed-budget setting, we formally define the multi-objective optimization (MOO) problem for each time instance $i$ below:
\begin{subequations}

\begin{align}
        &Maximize\; \bigcup_{w_i^j \in W_i} C(w_i^j) \label{eq:first_obj} \\ 
        &Minimize\; \max_{w_i^j \in W_i}(count_i(w_i^j)) \label{eq:second_obj} \\ 
        &s.t.\; \sum_{j=1}^{|W_i|}d(w_i^j)\le K_i  \label{eq:constraint} 
\end{align}
\end{subequations}
Equation \ref{eq:first_obj} maximizes the coverage of the selected workers while Equation \ref{eq:second_obj} minimizes the highest activation count across all workers present at the time.
The constraint in Equation~\ref{eq:constraint} ensures the number of selected workers does not exceed the budget $K_i$ at each time instance.

Rather than coming up with heuristics to sort the workers according to two objectives, we adopt a widely used approach, i.e., \underline{n}ondominated \underline{s}orting \underline{g}enetic \underline{a}lgorithm (NSGA)~\cite{srinivas1994muiltiobjective}, to solve the MOO formulation for each time instance.  Intuitively, nondominated sorting is to maintain stable nondominated fronts (i.e., subpopulations of good individuals) in a multi-dimensional space, where each dimension corresponds to an objective.  A nondominated front, also referred to as Pareto optimal, is a solution where none of the objective functions can be improved in value without degrading other objective values.   The advantage of genetic algorithms is that they simultaneously deal with a set of possible solutions i.e., population, which enables us to find several members of the Pareto optimal set in a single run of the algorithm. We outline our solution based on the NSGA algorithm in Algorithm~\ref{alg:ga}.

\begin{algorithm} [H]
\caption{\sc NSGA Algorithm}
\small
\begin{algorithmic}[1]
\STATE Population $P(t)\leftarrow$ RandomInit, $t \leftarrow 0$
\STATE While $t < maxgen$
\STATE \hspace{10pt} $P(t)' \leftarrow$ Select nondominated fronts $\{P(t)\}$, ranked by Eqs. \ref{eq:first_obj} and \ref{eq:second_obj}
\STATE \hspace{10pt} $P(t)'' \leftarrow$ Mutation $\{P(t)' \cup$ Crossover $\{P(t)'\}\}$ \label{line:cm}
\STATE \hspace{10pt} $P(t+1) \leftarrow P(t) \cup P(t)''$
\STATE \hspace{10pt} $t \leftarrow t + 1$
\STATE Select best solution from $P(t)$, ranked by Eq. \ref{eq:priori_ap} \label{line:priori_ap}
\end{algorithmic}
\label{alg:ga}
\end{algorithm}
  
The results of NSGA, at the end of \texttt{while} loop, include a set of nondominated fronts. Subsequently, in Line~\ref{line:priori_ap} we select the best individual solution based on a weighted sum of objective values: 
\begin{equation}
\alpha |\bigcup C(w_i^j)|/|T_i| - (1-\alpha)\max(count(w_i^j))/Q 
\label{eq:priori_ap}
\end{equation}
In Equation~\ref{eq:priori_ap}, $\alpha$ is a linear coefficient, $0 < \alpha < 1$, to specify the weight for each objective. The higher $\alpha$, the more important the objective of the Equation \ref{eq:first_obj} in comparison to that of Equation \ref{eq:second_obj}. The minus sign indicates the minimization objective in Equation \ref{eq:second_obj}. Both objective functions are normalized by the total number of tasks $|T_i|$ and the total number of time instances $Q$, respectively.
In our experiments, we adopted NSGA-II version~\cite{deb2002fast} for implementing Algorithm~\ref{alg:ga} and set $maxgen = 50,000$.

\subsection{Dynamic Budget $d\MCMC$}
\label{sec:ol_dynamic}
In the dynamic-budget setting, the multi-objective optimization (MOO) formulation is similar to Equation~\ref{eq:first_obj}, \ref{eq:second_obj} but the total budget is constrained over all time periods. Therefore, the constraint \ref{eq:constraint} is replaced by the following constraint: 
\begin{equation}
\sum_{i}\sum_{j=1}^{|W_i|}d(w_i^j)\le K 
\label{eq:ommdcontraint}
\end{equation}
$$ $$
In the online setting, we need to simultaneously consider the task coverage and the number of activations of the candidate worker, in order to optimize both objectives.
As a result, we modify the adaptive strategy in Section~\ref{sec:adaptive_budget} and define the gain $\lambda_l$ of the current worker $w^j_i$ in (\ref{eq:deltaGain}) to be a linear combination of the number previous activations and his priority: 

\begin{equation}
\lambda_l = \alpha. priority(w^j_i) / |T_i| - (1-\alpha)count(w^j_i)/Q
\label{eq:new_lambda}
\end{equation}
 
In equation \ref{eq:new_lambda},  $priority(w_i^j)$ and $ count(w_i^j)$ are respectively the priority of the worker calculated by any previously proposed local heuristic, and the number of times that worker was selected.  The coefficient $\alpha$ can be varied to balance the importance of the overloading and the priority. 

\section{Distance-based Task Utility}
\label{sec:task_value}
\begin{figure}[!htb]\centering
  \includegraphics[width=0.7\textwidth]{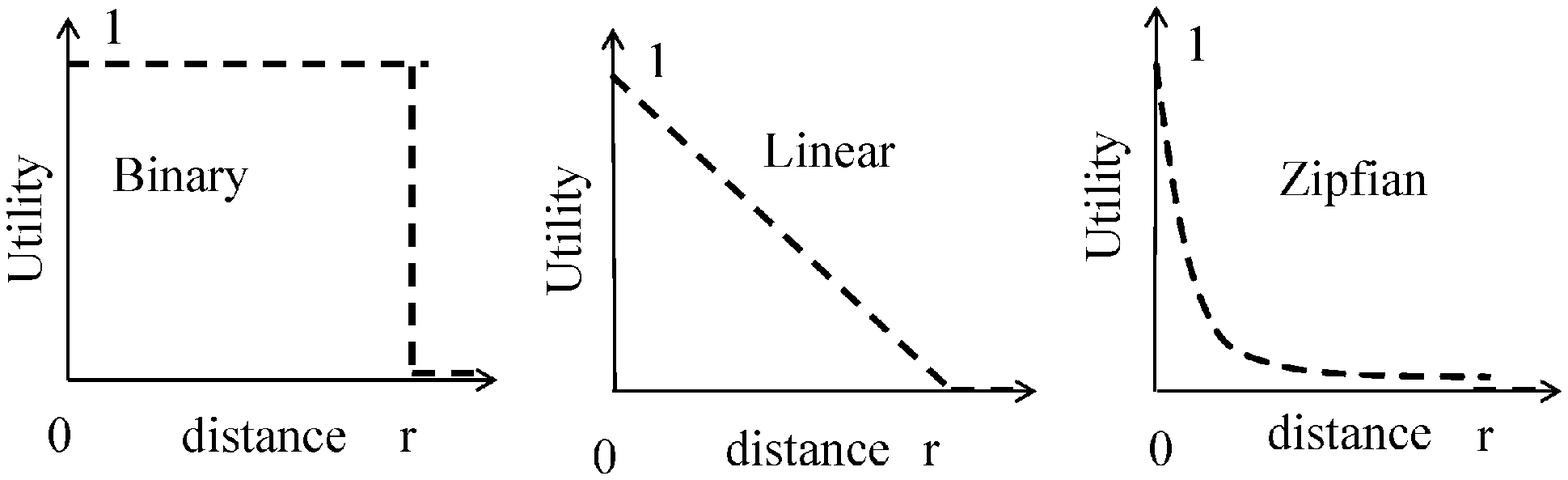}
  \caption{Distance-based utility functions.}
  \label{fig:task_model}
\end{figure}

Thus far, our goal is to maximize the number of assigned tasks, assuming assigning to any worker within the task region is equivalent.  However, in practice an assignment of a task to a nearby worker may yield higher \textit{utility} than that of a farther worker~\cite{miao2016balancing}.
Thus, in this section we aim to generalize  the binary-utility model to a distance-based-utility variant, i.e., maximizing the \emph{utility} of covered tasks. We assume the utility of worker $w$'s response to task $t$ is a function of the spatial distance between them: $utility(t,w)=f(dist(t,w))$.
And $f$ is a decreasing function of worker-task distance. Intuitively, the utility is at the highest when the worker is co-located with the task and decreases as the worker-task distance increases. The utility is zero if the distance is larger than task radius. We consider three cases depicted in Figure \ref{fig:task_model}: (i) \emph{Binary}, where utility has value 1/0 (ii) \emph{Linear}, where utility decreases linearly with the worker-task distance and (iii) \emph{Zipf}, where utility follows \emph{Zipfian} distribution with skewness parameter $s$. The higher the value of $s$, the faster utility drops.
This extension can be incorporated into all the previously developed algorithms. Specifically, 
Algorithm \ref{alg:basic} (Line \ref{line:select_w}) now chooses the worker that maximizes utility increase at each stage.
\begin{equation}
\label{eq:utility_increment}
priority(w_i^j)=\sum\limits_{t_i^k\in C(w_i^j)\cap R}f(dist(t_i^k, w_i^j))
\end{equation}

With the temporal heuristic, Equation \ref{eq_ctdp} becomes:
\begin{equation}
priority(w_i^j)=\sum_{t_i^k \in C(w_i^j)\cap R}\frac{f(dist(t_i^k,w_i^j))}{t_i^k.(\delta + s)-i}
\label{eq_ctdp_utility}
\end{equation}

In the same fashion, with the spatial heuristic, Equation \ref{eq:weight_re} becomes:
\begin{equation}
\label{eq:weight_re_utility}
priority(w_i^j)=\sum_{t_i^k \in C(w_i^j)\cap R}\frac{f(dist(t_i^k,w_i^j))}{1 + RE(t_i^k)}
\end{equation}

With the adaptive budget strategies, the gain of a candidate worker is adapted similarly.


\section{Performance Evaluation}
\label{sec:exp}


\subsection{Experimental Methodology}
\label{sec:methodology}

We adopted real-world datasets from location-based applications, summarized in Table~\ref{tab:real_datasets}, to emulate spatial crowdsourcing (SC) workers and tasks. We consider Gowalla, Foursquare users as SC workers and the venues as tasks.
The Gowalla dataset contains check-ins for 224 days in 2010, including more than 100,000 spots (e.g., restaurants), within the state of California.
By considering each day as a unit time period, all the users who checked in during a day are available workers for that time period in our setting.  
The Foursquare dataset contains the check-in history of 45,138 users to 89,968 venues over 384 hours in Pittsburgh, Pennsylvania.  We considered each hour as a unit time period for this dataset.  
%
%

\begin{table}
\begin{center}
\footnotesize
\begin{tabular}{ l | c |  c |  c | r}
\hline
\textbf{Name} & \#Tasks & \#Workers & MTD~\footnote{MTD: Mean Travel Distance \cite{to2014framework}} & $|s_i|$ \tn
\hline
Foursquare & 89,968 & 45,138 (90/$km^2$) & 16.6km & 1 hour \tn
\hline
Gowalla & 151,075 & 6,160 (35/$km^2$) & 3.6km & 1 day  \tn
\hline
\end{tabular}
\caption{Summaries of real-world datasets.}
\label{tab:real_datasets}
\end{center}
\end{table}

We generated a range of datasets with SCAWG toolbox \cite{to2016scawg} by utilizing real-world worker/task spatial distributions and varying their arrival rate. 
We generated worker count following COSINE (default) and POISSON distributions with mean = 50 and set default value of task count per time period to be constant, i.e., 1000. We denote Go-POISSON a dataset that uses Gowalla for the spatial distribution and POISSON for the worker arrival rate.




In all of our experiments, for Gowalla dataset,  we varied the total number of time periods $Q\in$ \{7, 14, {\bf 28}, 56\} and the budget $K\in$ \{56, 112, 224, {\bf 448}, 896, 1288\}. For Foursquare, $Q\in$ \{{\bf 24}, 48, 72, 96\} and $K\in$ \{24, 48, 96, {\bf 192},..., 1536\} because we modeled a time period as one hour.
  The task duration $\delta$ was randomly chosen from 1 to $Q$ and the task radius $r\in$ \{1, 2, 3, 4, {\bf 5}, 6, 7, 8, 9, 10\} km. The choices of $r$ and $\delta$ values are defined by the CHRS experts. Default values are shown in boldface.
Finally, experiments were run on an Intel(R) Core(TM)i7-2600 CPU @ 3.40 GHz with 8 GB of RAM.

\subsection{Experimental Results}
In the following we evaluate our solutions in terms of the number of covered tasks, i.e., \emph{task coverage}. We first show the performance of offline solutions (Section~\ref{sec:exp_offline}). Then we present the results for the online scenario, including local heuristics, adaptive budget strategy, and workload heuristic (Section~\ref{sec:exp_heuristics}). We next show the results of distance-based utility and worker overloading (Section \ref{sec:non-binary}), followed by runtime measurements (Section \ref{sec:runtim}).


\subsubsection{Optimal Solutions for Offline Setting}
\label{sec:exp_offline}

\begin{figure*}[ht]
	\centering
	\subfigure[Vary $K$, Go-COSINE]{\label{fig:go_cosine_fd_vary_b}\includegraphics[width=.42\textwidth]{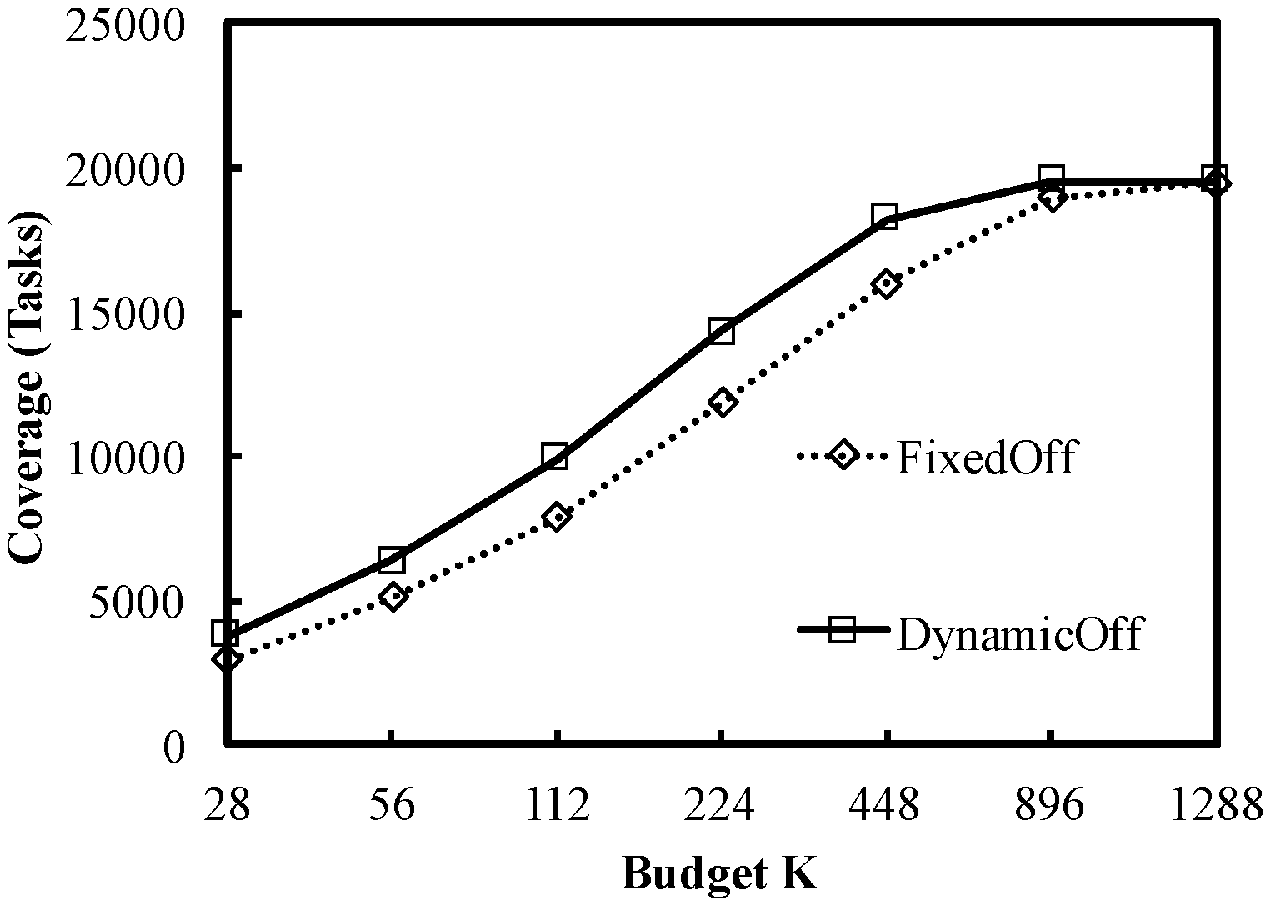}}
	\hspace{30pt}
	\subfigure[Vary $r$, Go-COSINE]{\label{fig:go_cosine_fd_vary_r}\includegraphics[width=.42\textwidth]{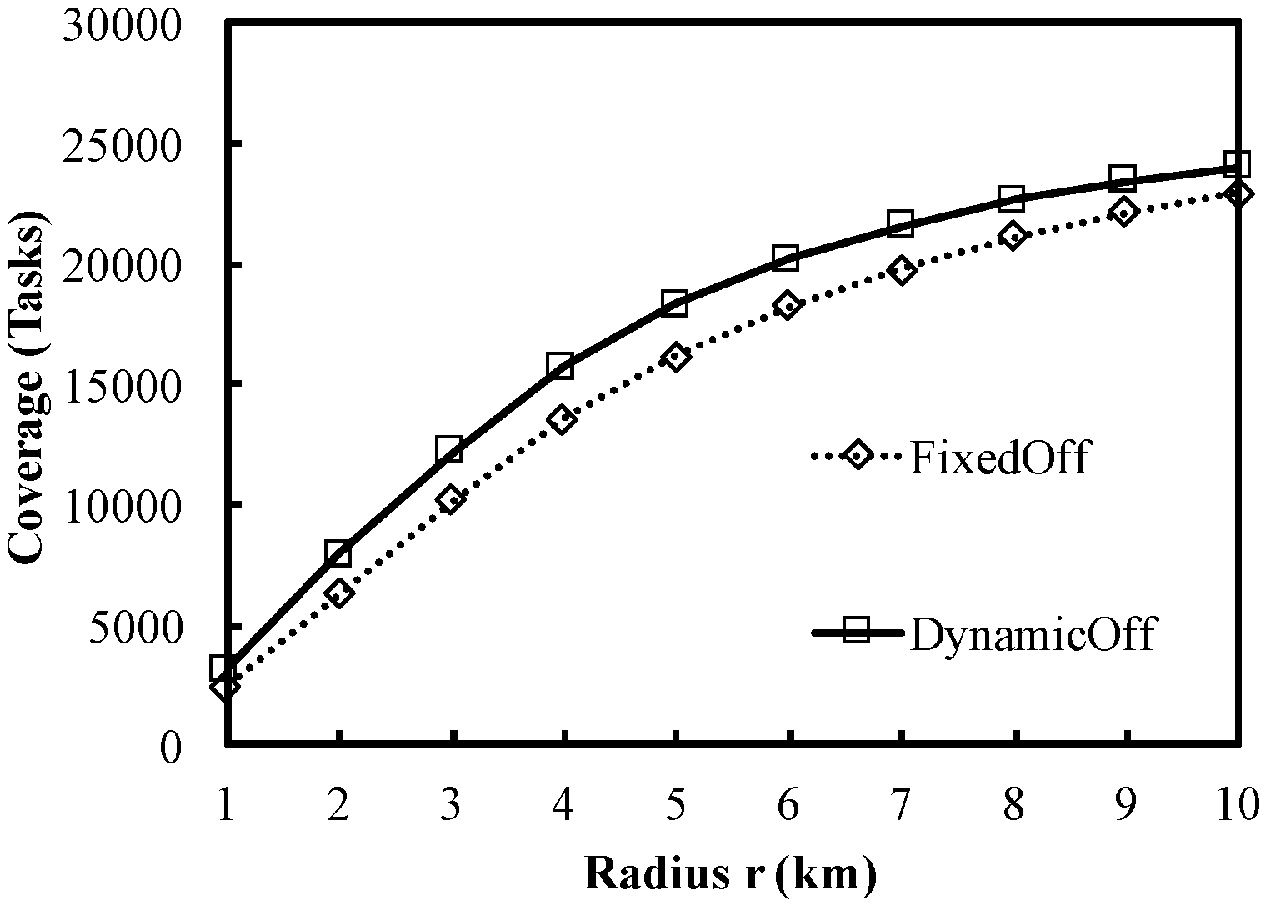}}
	
	\subfigure[Vary $K$, Go-POISSON]{\label{fig:go_poisson_fd_vary_b}\includegraphics[width=.42\textwidth]{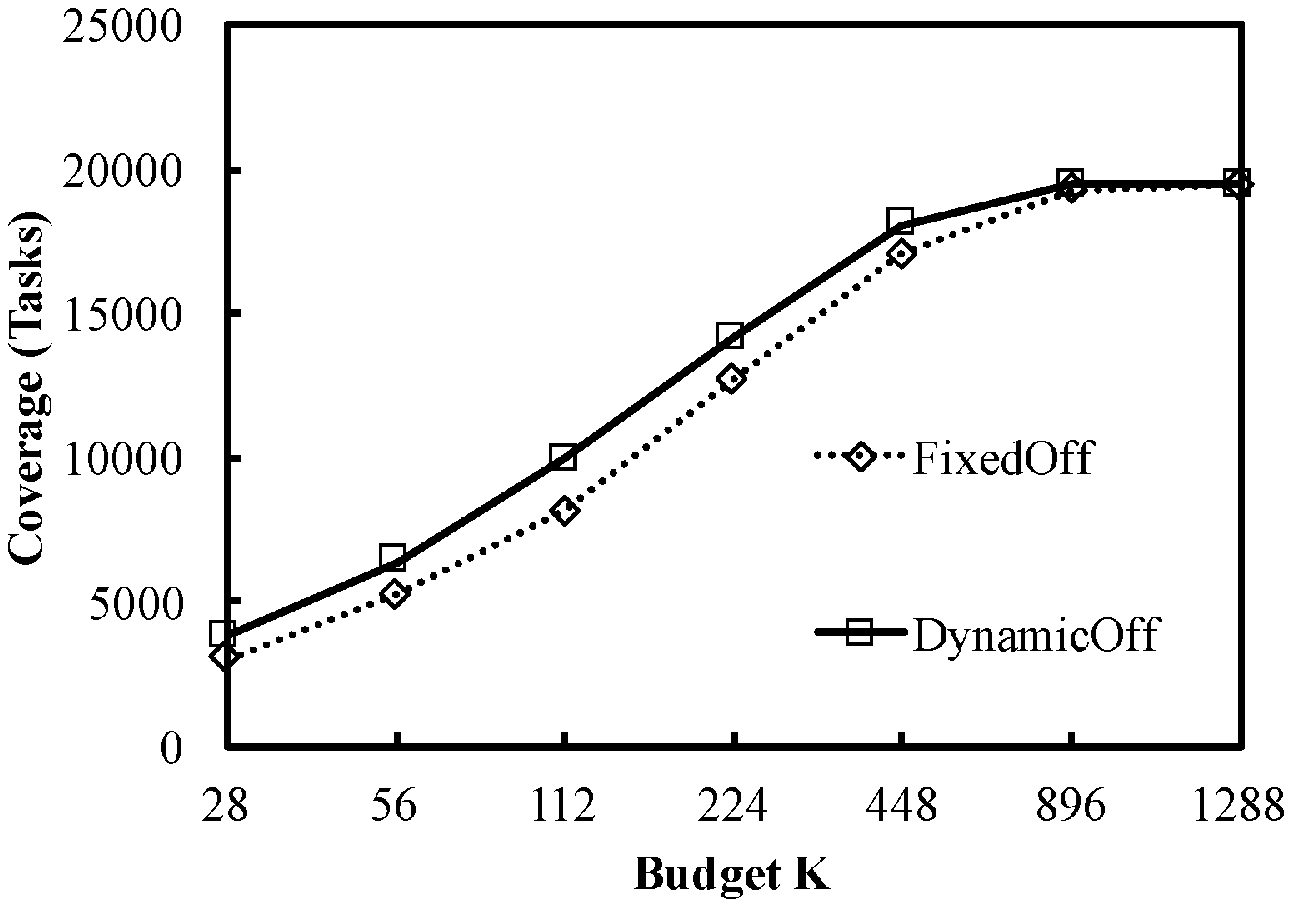}}
	\hspace{30pt}
	\subfigure[Vary $r$, Go-POISSON]{\label{fig:go_poisson_fd_vary_r}\includegraphics[width=.42\textwidth]{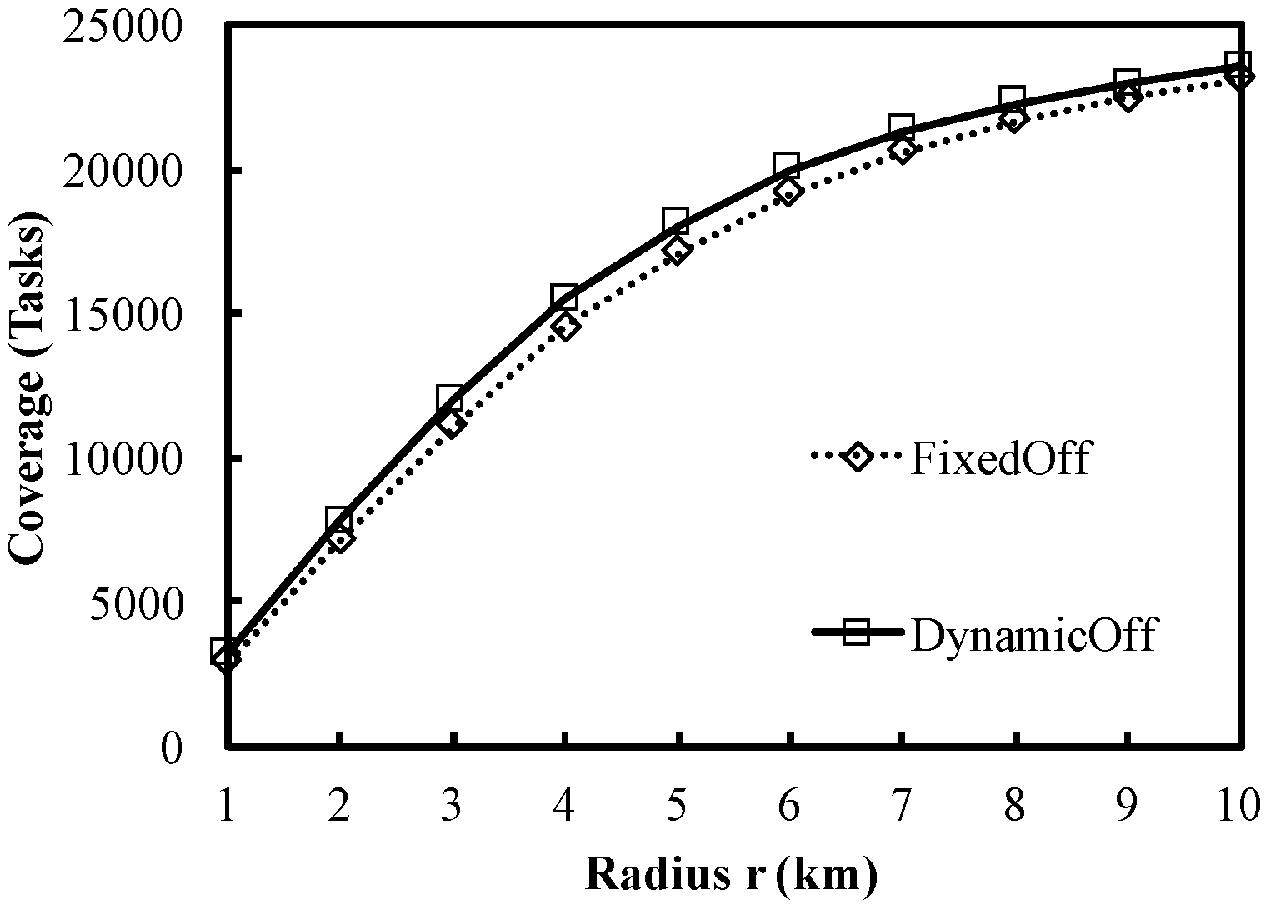}}
	\caption{Performance of offline solutions with Go-COSINE and Go-POISSON.}
	\label{fig:go_fix_dynamic}
\end{figure*}

We implemented the offline solutions for the two problem variants, \emph{$f{\MCMC}$} and \emph{$d{\MCMC}$} (Section~\ref{sec:hardness}), using integer linear programming. These algorithms provide optimal results, which are used as the upper-bounds of the online algorithms.
Figure \ref{fig:go_cosine_fd_vary_b} illustrates the results for Go-POISSON by varying the total budget $K$. As expected, higher budget yields higher coverage. Also, \emph{DynamicOff} yields higher coverage than \emph{FixedOff} as \emph{FixedOff} is the special case of \emph{DynamicOff}. However, the higher the budget, the smaller the performance gap between them. 
This effect can be explained by the diminishing return property of the max cover problem. That is, the more workers are selected, the smaller gain of each selected worker.
We also evaluated the offline solutions by varying task radius $r$ (Figure \ref{fig:go_cosine_fd_vary_r}). Intuitively,
when $r$ increases, more workers are located within the task's spatial range. This means that a worker is eligible to perform more tasks, which yields higher task coverage.


Similar trends were observed for Go-POISSON as shown in Figures~\ref{fig:go_poisson_fd_vary_b} and~\ref{fig:go_poisson_fd_vary_r}. We observe a small difference between  \emph{FixedOff} and \emph{DynamicOff} for Go-POISSON in Figure \ref{fig:go_poisson_fd_vary_r}.  However, when the arrival rate has high variance, such as in Go-COSINE, \emph{DynamicOff} shows more coverage over \emph{FixedOff} in Figure \ref{fig:go_cosine_fd_vary_r}.
The reason is that \emph{FixedOff} uses a constant budget to the time periods with high spikes while \emph{DynamicOff} can allocate more budget to those time periods to cover more tasks.

\subsubsection{Local Heuristics and Budget Allocations for Online Setting} 
\label{sec:exp_heuristics}
\begin{figure*}[ht]
		\centering
		\subfigure[Vary $K$, Go-COSINE]{\label{fig:go_cosine_f_vary_b}\includegraphics[width=.42\textwidth]{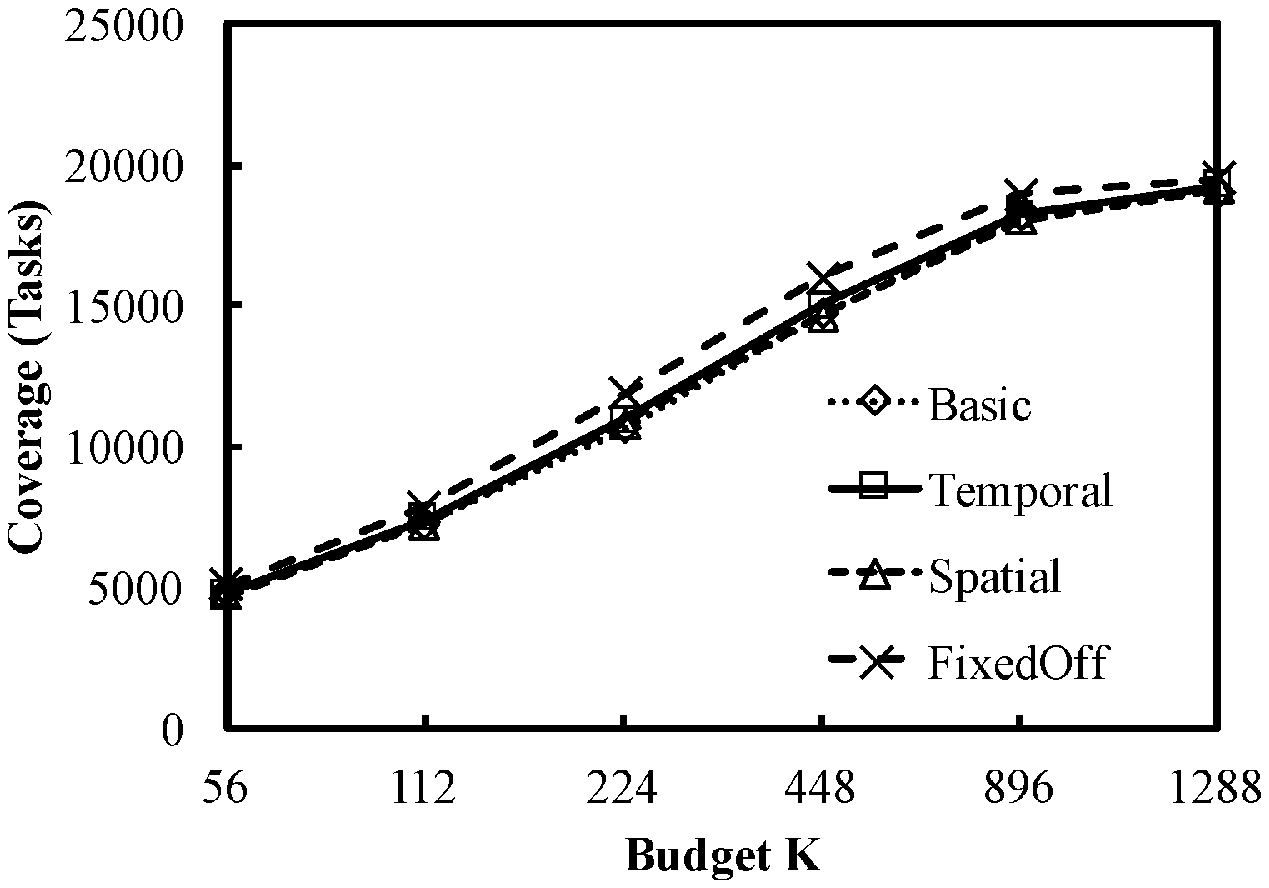}}
		\hspace{30pt}
		\subfigure[Vary $r$, Go-COSINE]{\label{fig:go_cosine_f_vary_r}\includegraphics[width=.42\textwidth]{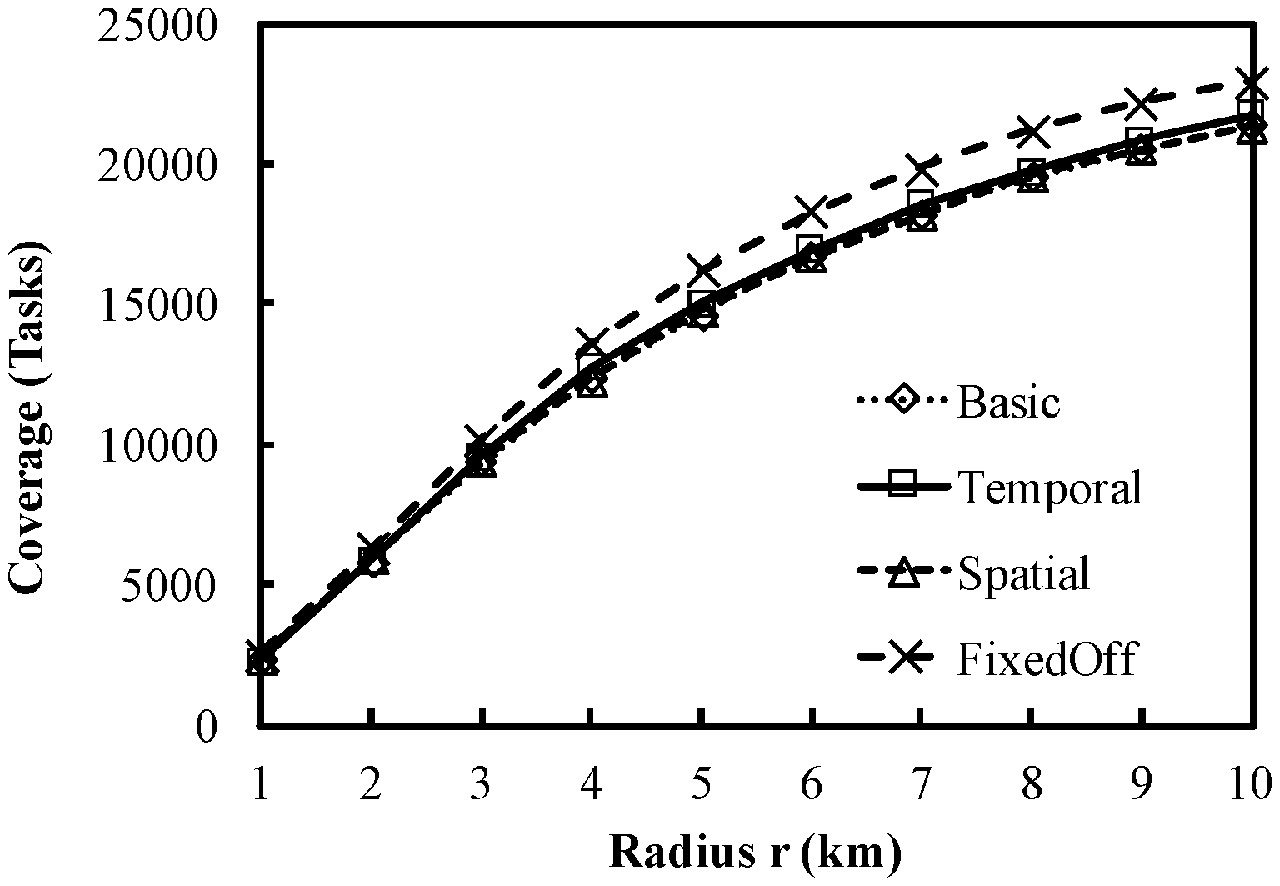}}
		\caption{Performance of local heuristics in the fixed-budget scenario.}
		\label{fig:local_algorithms}
	\end{figure*}

\textbf{\\The Performance of Heuristics:} We evaluate the performance of the online heuristics for \emph{$f{\MCMC}$}  from Section \ref{sec:fixed}, \emph{Basic}, \emph{Spatial} and \emph{Temporal}. Figures \ref{fig:go_cosine_f_vary_b} shows the improvements of \emph{Spatial} and \emph{Temporal} over \emph{Basic} on Go-COSINE. When the budget is high, we observe that the simple heuristic \emph{Basic} already obtains results close to the optimal solution. This is because most workers are selected. Furthermore, \emph{Spatial} and \emph{Temporal} yield 2\% and 4\% higher coverage than \emph{Basic} (at $K=224$) and their performance converges as $K$ increases.
Similar trends can be observed when increasing the task radius $r$ in Figure \ref{fig:go_cosine_f_vary_r}.


\begin{figure*}[ht]
		\centering
		\subfigure[Vary $K$, Go-COSINE]{\label{fig:baseline_vary_k}\includegraphics[width=.42\textwidth]{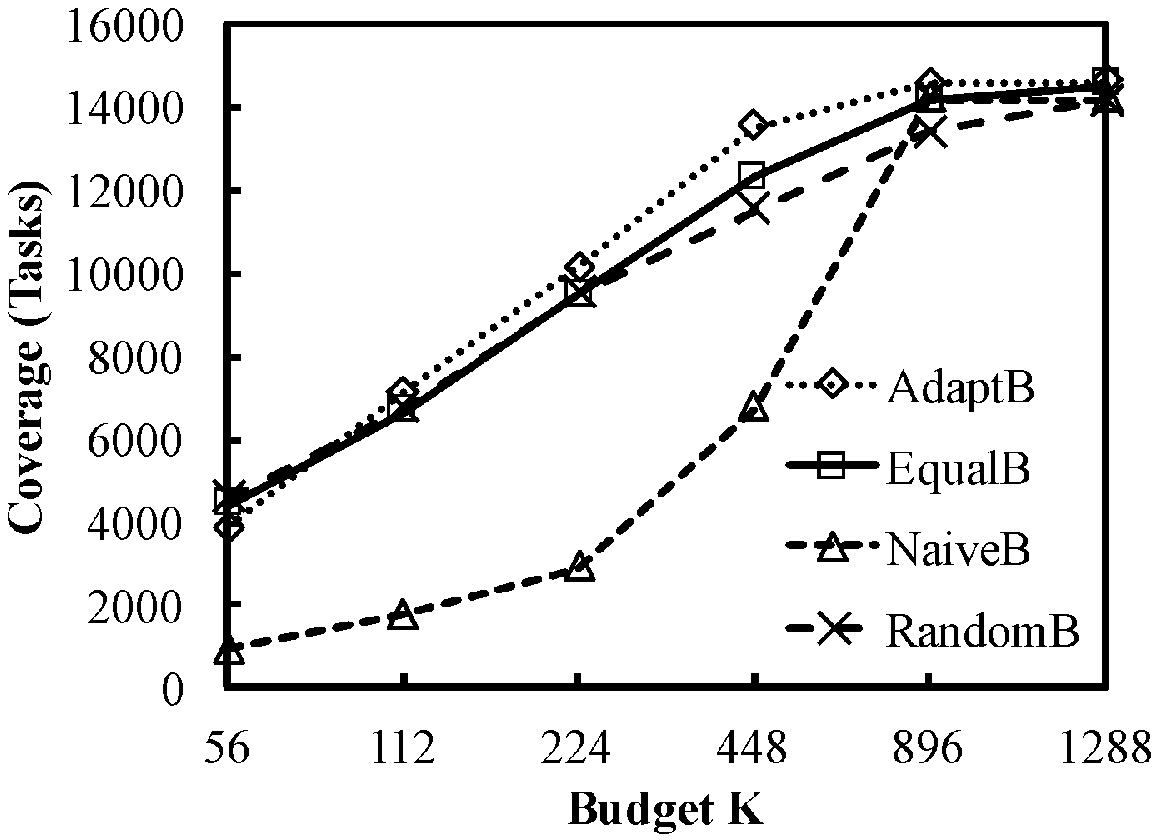}}
		\hspace{30pt}
		\subfigure[Vary $K$, Go-POISSON]{\label{fig:baseline_vary_k2}\includegraphics[width=.42\textwidth]{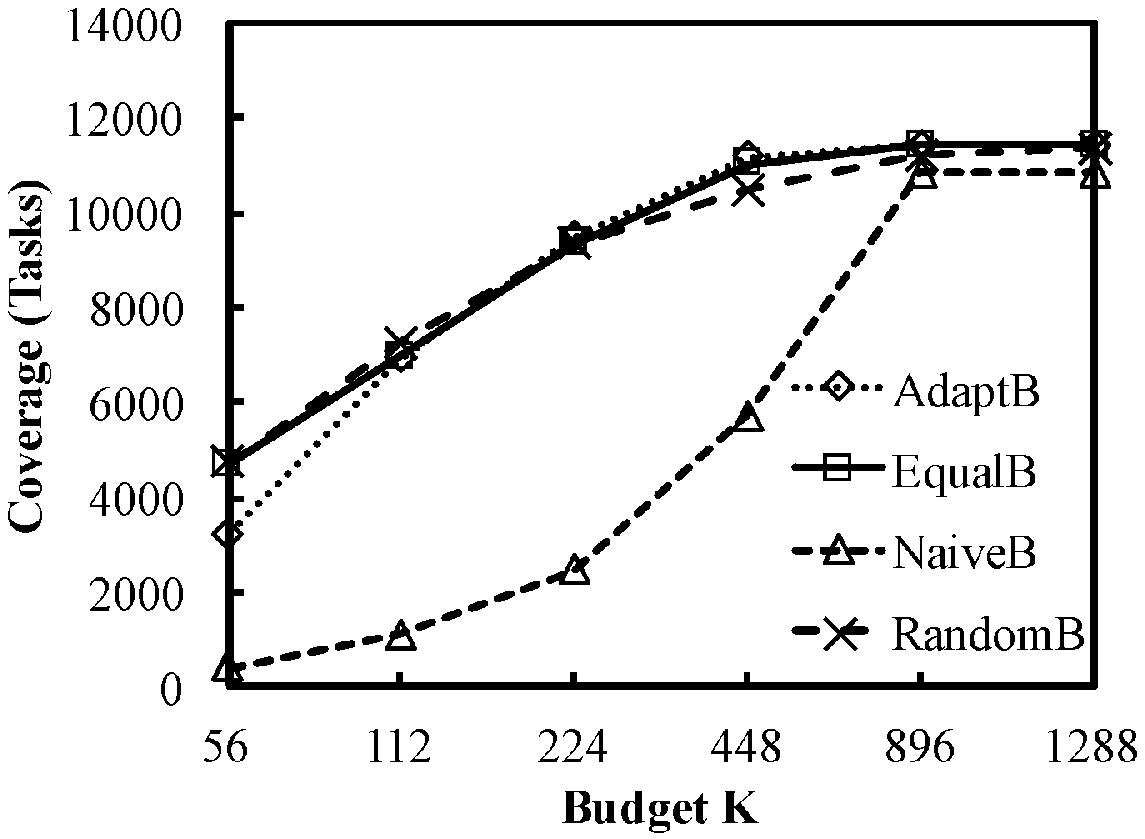}}

		\caption{Performance of adaptive budget allocation strategies.}
		\label{fig:baseline}
	\end{figure*}

	\textbf{Adaptive Budget Allocation Strategy:} We evaluate the performance of the adaptive budget allocation strategy in Section \ref{sec:adaptive_budget} by comparing with three baseline strategies inspired by a few previous studies, namely, \textit{Equal} \cite{to2016real,tran2013efficient}, \textit{Random} \cite{tran2013efficient} and \textit{Naive} \cite{kazemi2013geotrucrowd,ji2016urban,zhang2015event}. In \textit{Equal} budget strategy, the total budget is allocated equally to the time intervals.  In \textit{Random} budget strategy, a random positive number $k_i$ is generated for each time interval $i$ and then each time interval $i$ is given a budget $K_i = K* \frac{k_i}{\sum_{1}^{Q}k_i}$. In \textit{Naive} budget strategy, there is no particular limitation for the budget of each time interval, the budget is used until no more worker is available or entire budget is exhausted.
	We use local heuristic \textit{Basic} to compare the performances of budget strategies. Figure \ref{fig:baseline} shows the results of the budget strategies when varying the total budget $K$. \textit{AdaptB} is shown to be the best in coverage. \textit{EqualB} and \textit{RandomB} do not perform as well as \textit{AdaptB}, as they lack an intelligent budget allocation strategy. \textit{NaiveB} performs poorly as it selects the workers on a first-come-first-serve basis without considering future time intervals. As the result, the total budget is quickly exhausted during the  first few time intervals. The difference between \textit{AdaptB} and the others is higher with GO-COSINE because it has more fluctuation in the number of workers over time. This shows that \textit{AdaptB} can quickly adapt to the dynamic arrivals of workers.  
	
\begin{figure*}[ht]
	\centering
	
	\subfigure[Vary $K$, Go-COSINE]{\label{fig:go_cosine_d_vary_b}\includegraphics[width=.42\textwidth]{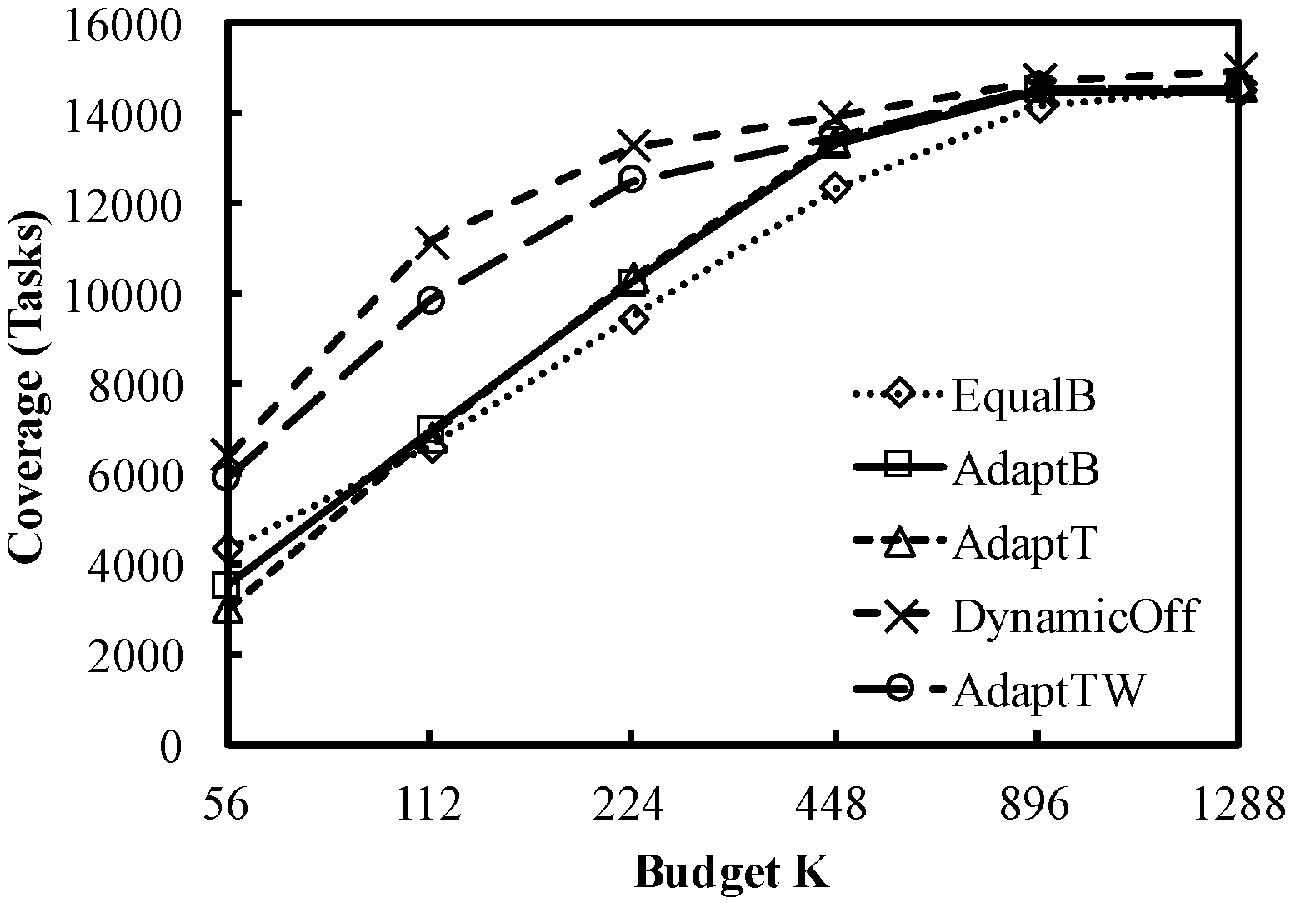}}
	\hspace{30pt}
	\subfigure[Vary $r$, Go-COSINE]{\label{fig:go_cosine_d_vary_r}\includegraphics[width=.42\textwidth]{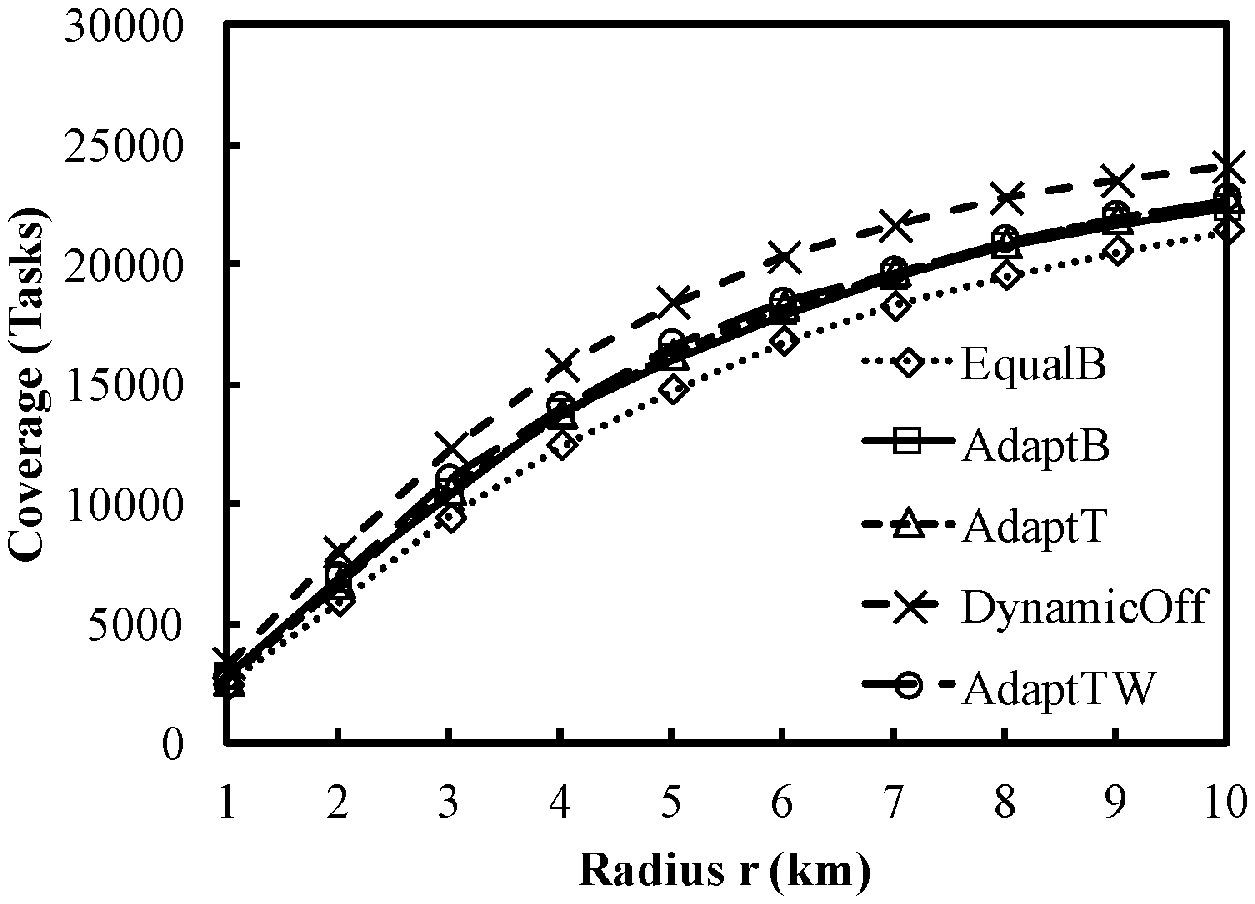}}
	
	\subfigure[Vary $K$, Go-POISSON]{\label{fig:go_poisson_d_vary_b}\includegraphics[width=.42\textwidth]{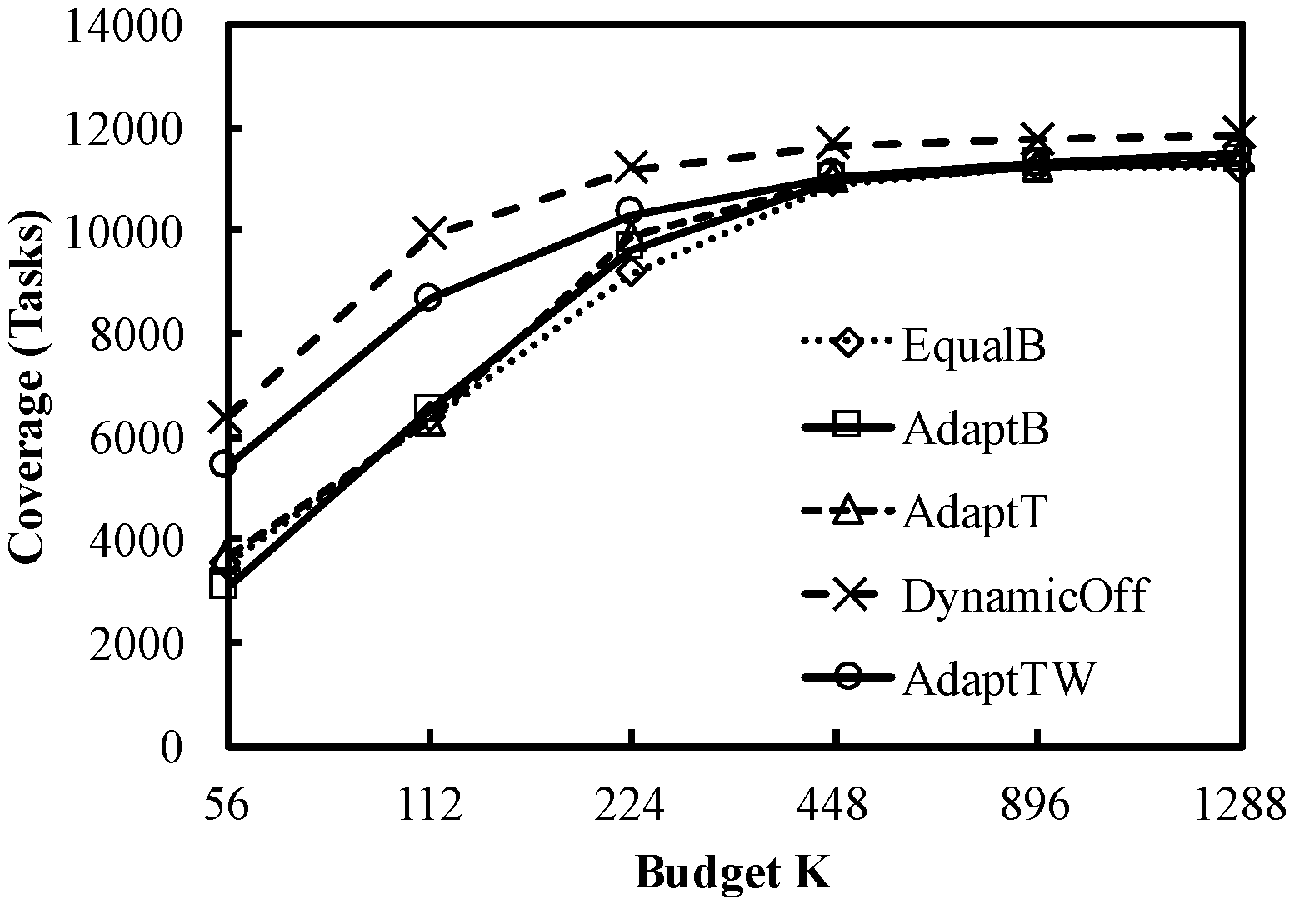}}
	\hspace{30pt}
	\subfigure[Vary $r$, Go-POISSON]{\label{fig:go_poisson_d_vary_r}\includegraphics[width=.42\textwidth]{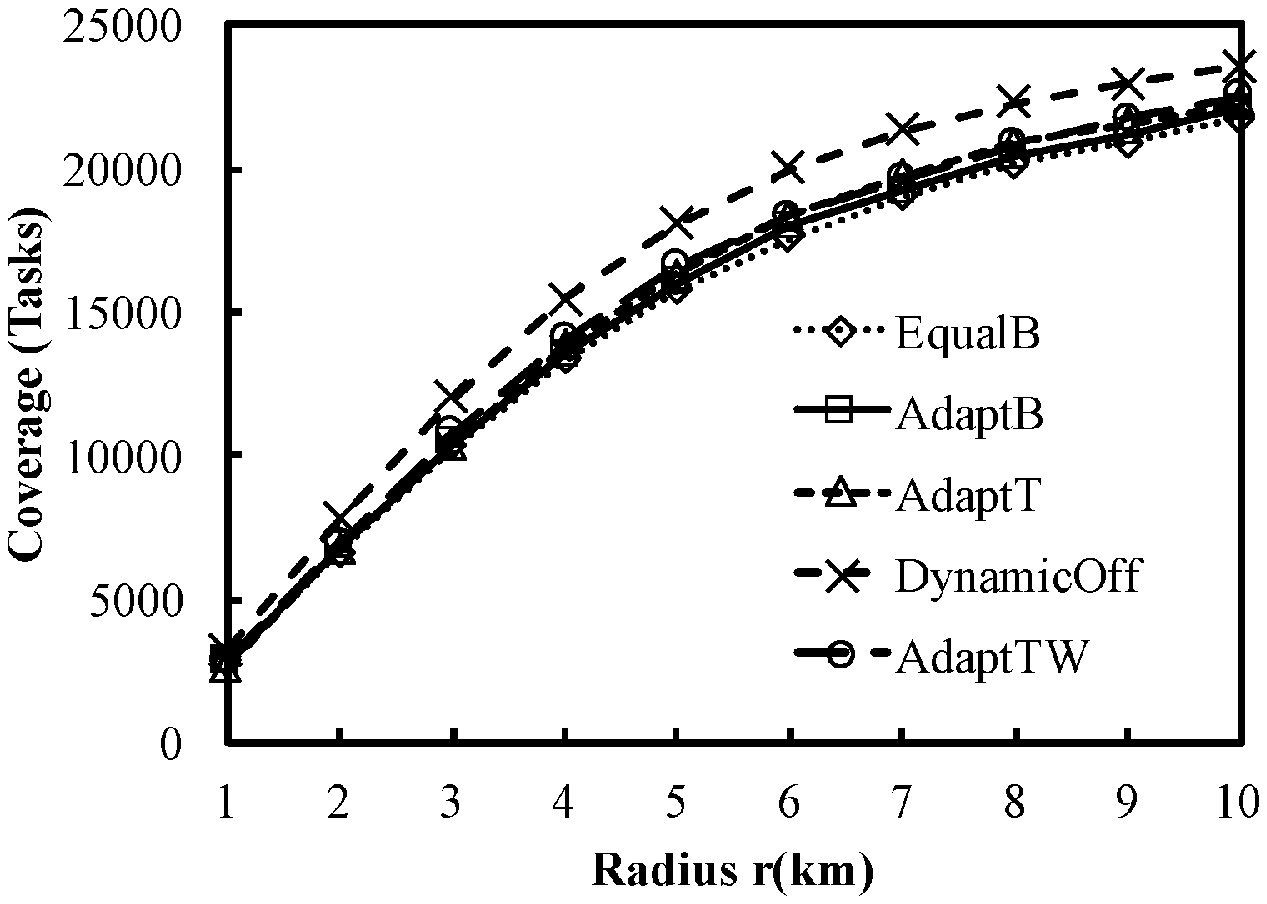}}
	
	
	\caption{Performance of adaptive budget allocation in the dynamic-budget scenario.}
	\label{fig:adapt}
\end{figure*}

\textbf{Historical Workload Improvement:} We also evaluate the performance of the adaptive budget allocation strategy applied with local heuristics, in which \textit{Temporal} is shown to perform better than \textit{Basic}. Figure \ref{fig:adapt} shows the results of \textit{EqualB}, \text{AdaptB}, \textit{AdaptT} and \textit{AdaptTW} (\textit{AdaptT} with historical workload improvement) when varying total budget $K$ and task radius $r$ on Go-COSINE and Go-POISSON datasets. We include  \textit{DynamicOff} as the optimal result for reference. As can be seen in that figure, with small budgets, \emph{AdaptTW}, which uses historical optimal workload as the baseline budget strategy, has higher coverage than the others and with $K = 56$, \textit{EqualB} performs better than \textit{AdaptB}, \textit{AdaptT}. The reason is with small budgets, \textit{AdaptB} and \textit{AdaptT} do not have enough contextual information.   With higher budget (K $\ge 448$), the adaptive algorithms perform better than \textit{EqualB} and their results are close to the optimal result. We observe similar results when varying $r$.  

\begin{figure*}[]
		\centering
		\subfigure[Go-COSINE]{\label{fig:confidence1}\includegraphics[width=.42\textwidth]{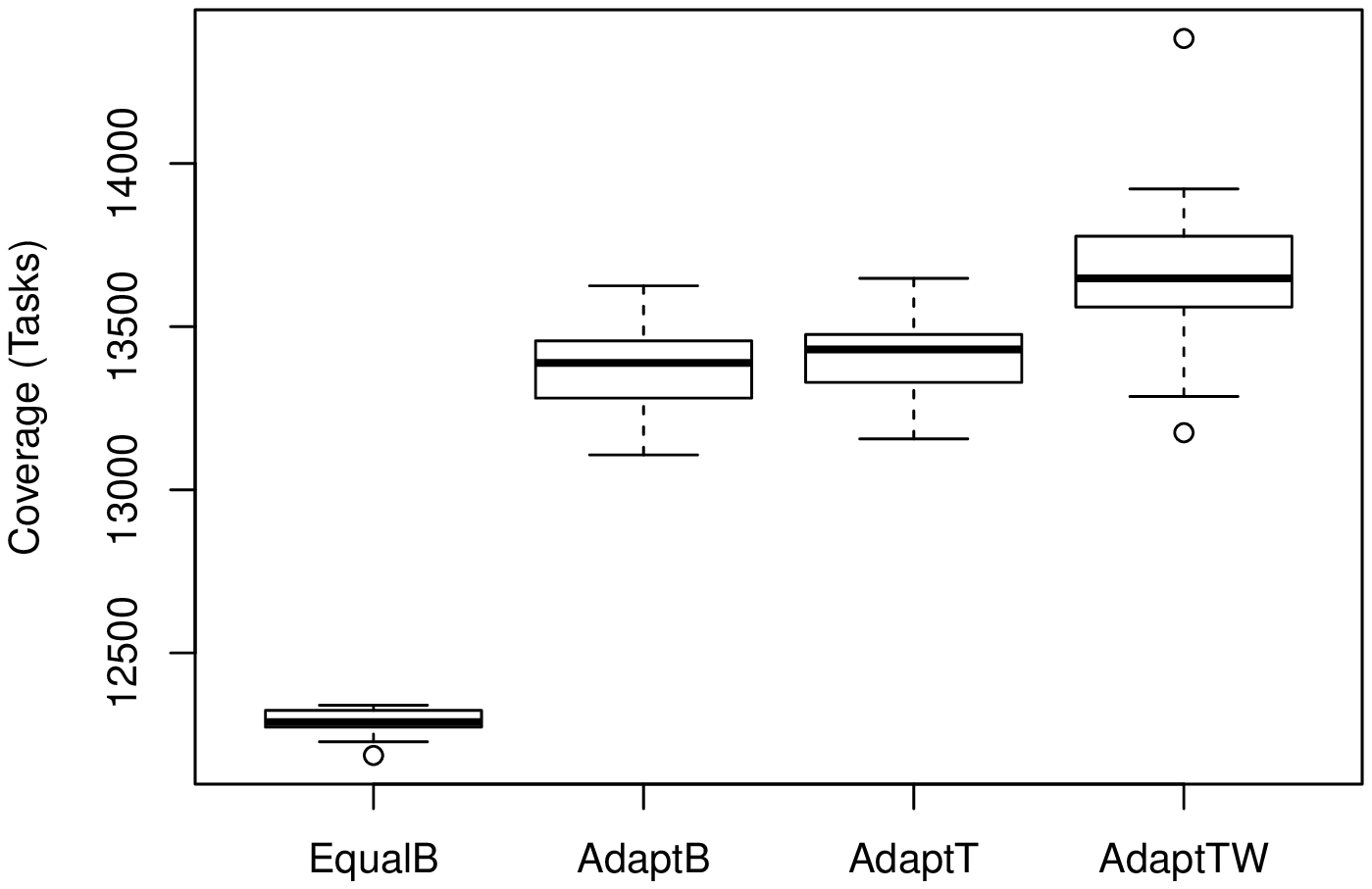}}
		\hspace{30pt}
		\subfigure[ Go-POISSON]{\label{fig:confidence2}\includegraphics[width=.42\textwidth]{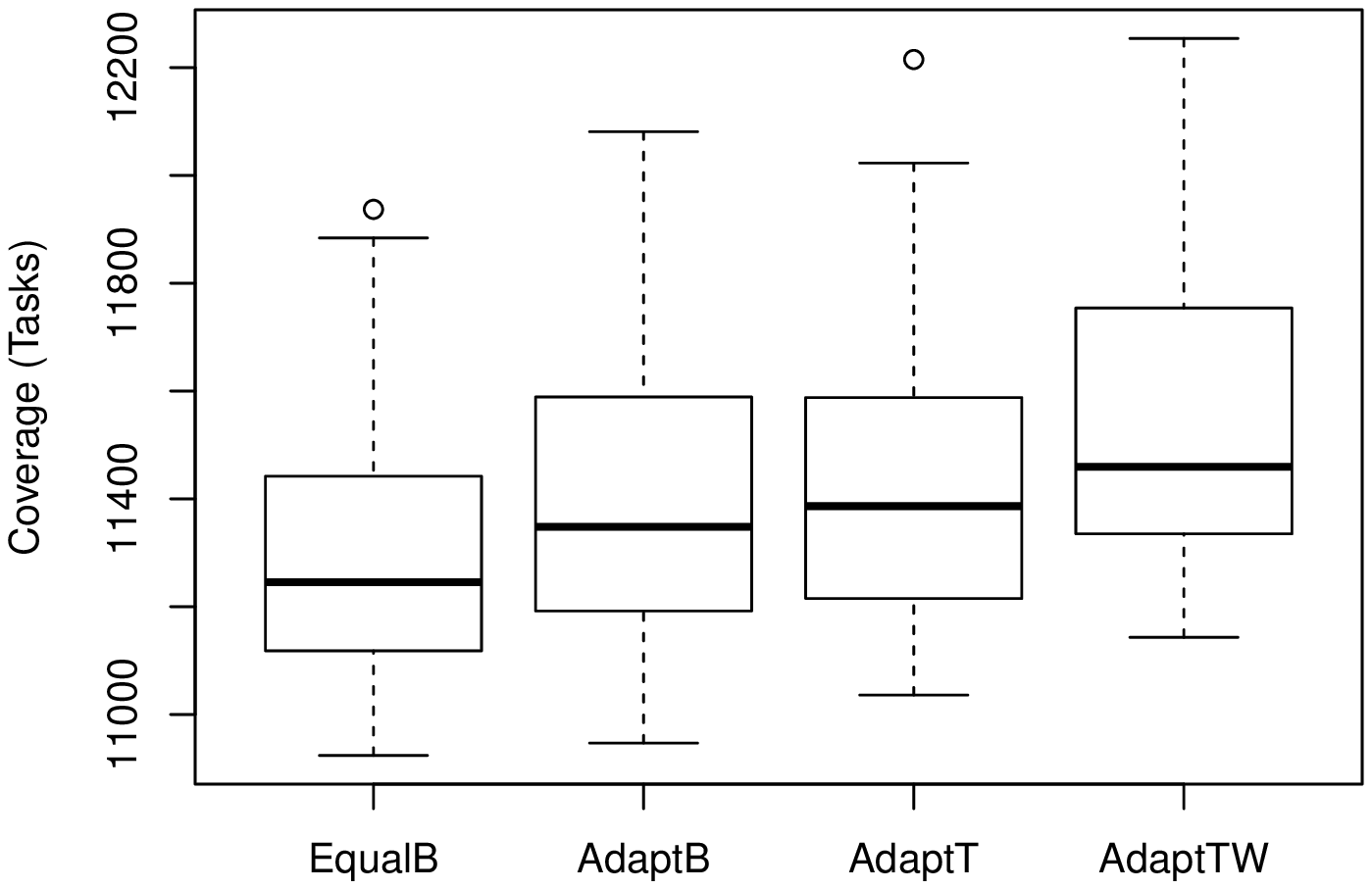}}
		
		\caption{Boxplots for various budget allocation strategies.}
		\label{fig:confidenceInterval}
	\end{figure*}

We further study the performance of various  budget allocation strategies by plotting the task coverage across multiple workloads using boxplots. Figure \ref{fig:confidenceInterval} shows the results of the techniques with the default parameter setting. As can be seen in the figure, the adaptive algorithms perform better than \textit{EqualB} especially with GO-COSINE dataset. While \textit{AdaptTW} has the highest median, minimum, maximum values, \textit{AdaptT} is the most stable method with the smallest difference between the minimum and the maximum values.  

\subsubsection{Distance-based Task Utility and Worker Overload}
\label{sec:non-binary}

\begin{figure*}[ht]
	\centering
	\subfigure[Task coverage (vary $K$)]{\label{fig:go_cosine_db_vary_b}\includegraphics[width=.42\textwidth]{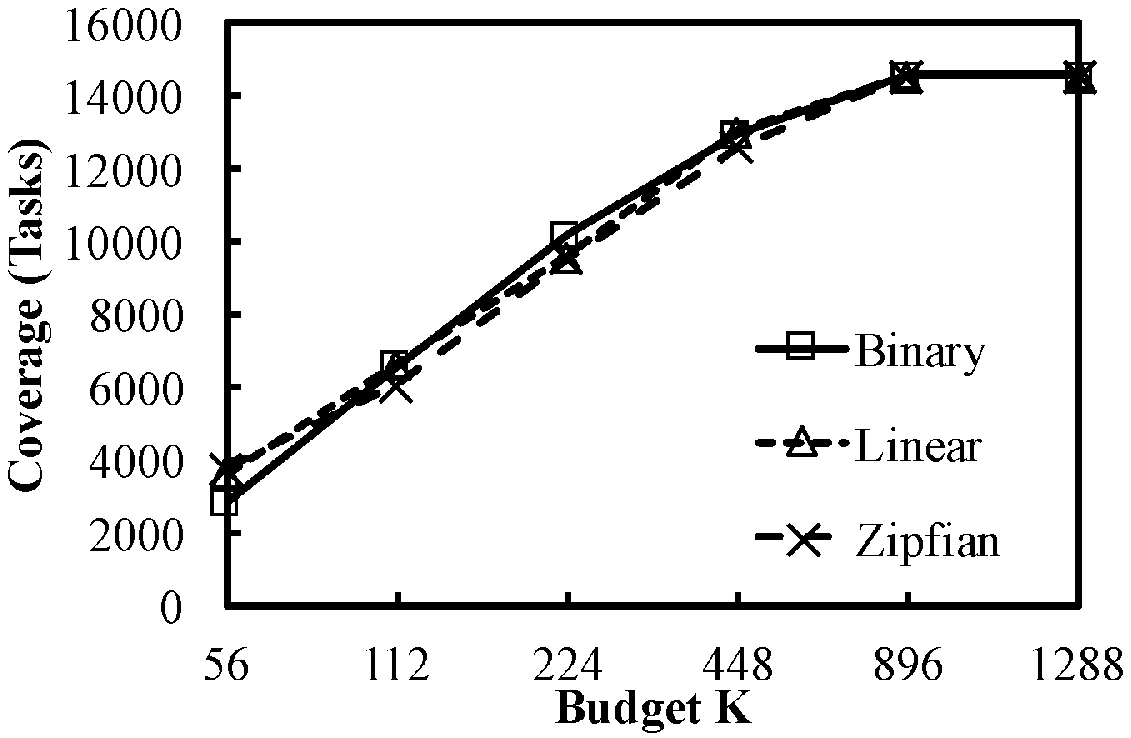}}
	\hspace{30pt}
	\subfigure[Overlapping ratio (vary $K$)]{\label{fig:go_cosine_db_vary_b2}\includegraphics[width=.42\textwidth]{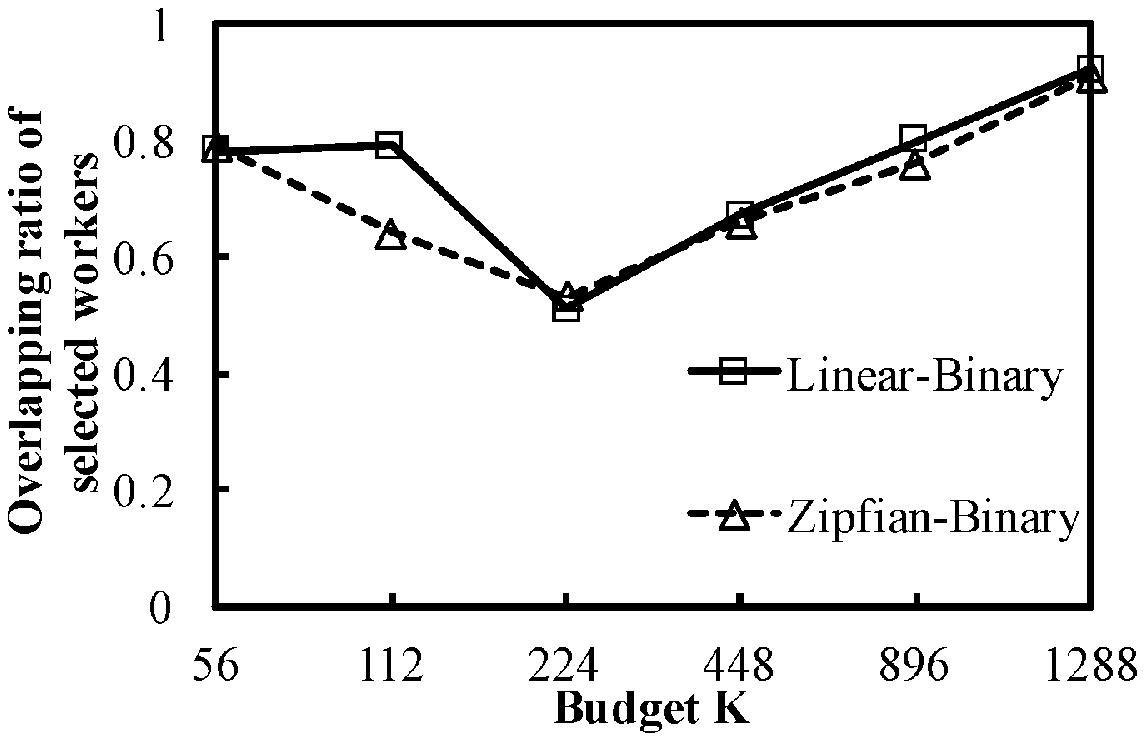}}
	
	\caption{Performance of \emph{AdaptT} on Go-COSINE with distance-based utility. The overlapping ratio indicates the percentages of workers that are selected in both binary and  the corresponding distance-based utility model.}
	\label{fig:db_workload}
\end{figure*}

\textbf{\\Distance-based Task Utility:}
We show the performance of \emph{AdaptT} under distance-based functions for task utility from Section~\ref{sec:task_value}. In Figure \ref{fig:go_cosine_db_vary_b}, we observe that the obtained task coverage in the cases of Linear and Zipfian are similar to the result in the binary-utility model. We also present the overlapping ratio of selected workers between distance-based utility and binary utility as shown in Figure~\ref{fig:go_cosine_db_vary_b2}. The figure shows that when total budget increases, initially, the ratio decreases because of more different workers can be selected and then ratio increases because when the budget is large enough, most workers are selected in both cases, resulting in large overlaps.

\begin{figure}[ht]
	\centering
	\subfigure[Task coverage (vary $\alpha$)]{\label{fig:go_cosine_ga_vary_a}\includegraphics[width=.42\textwidth]{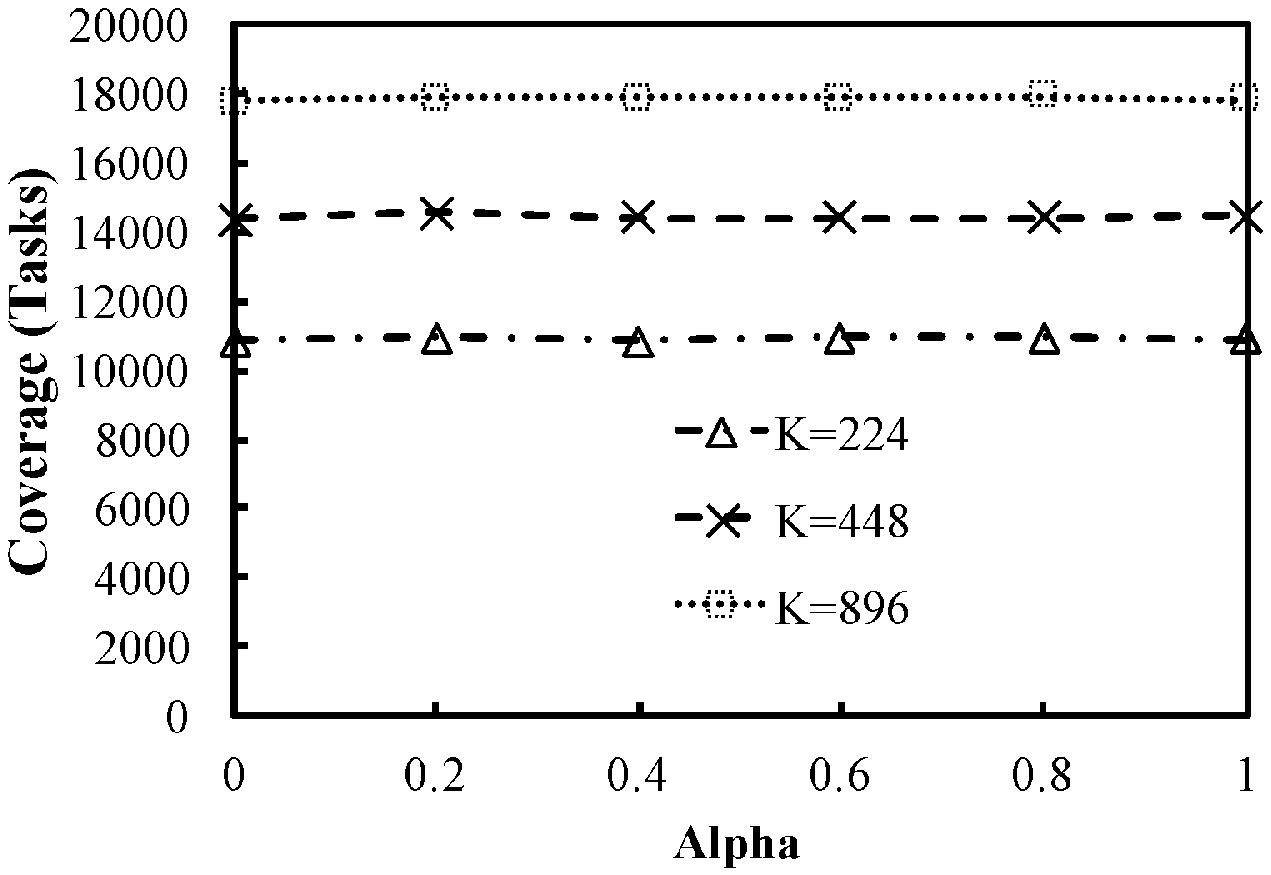}}
	\hspace{30pt}
	\subfigure[Average number of activations (vary $\alpha$)]{\label{fig:go_cosine_ga_vary_a2}\includegraphics[width=.42\textwidth]{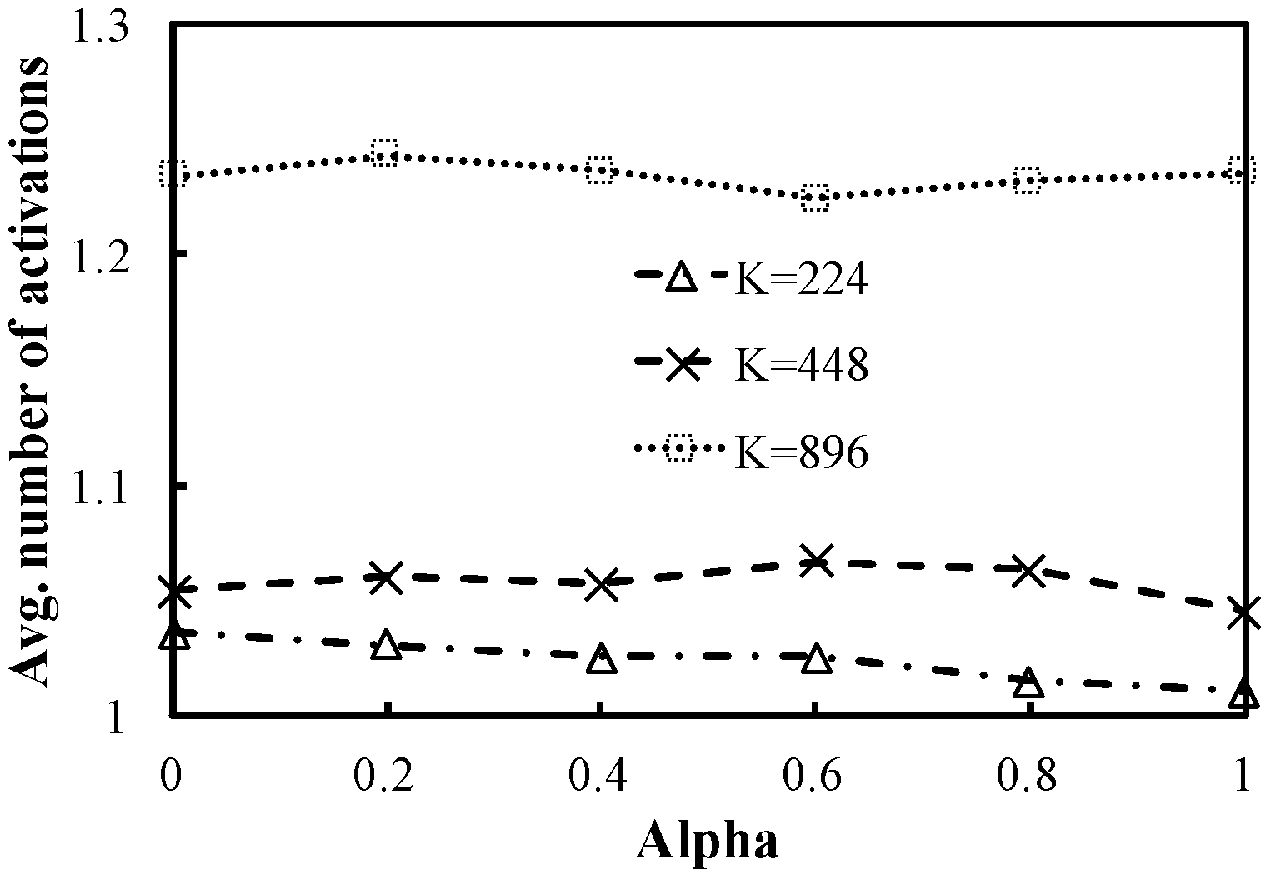}}
	
	\subfigure[Task coverage (vary $\alpha$)]{\label{fig:go_cosine_ol_vary_a}\includegraphics[width=.42\textwidth]{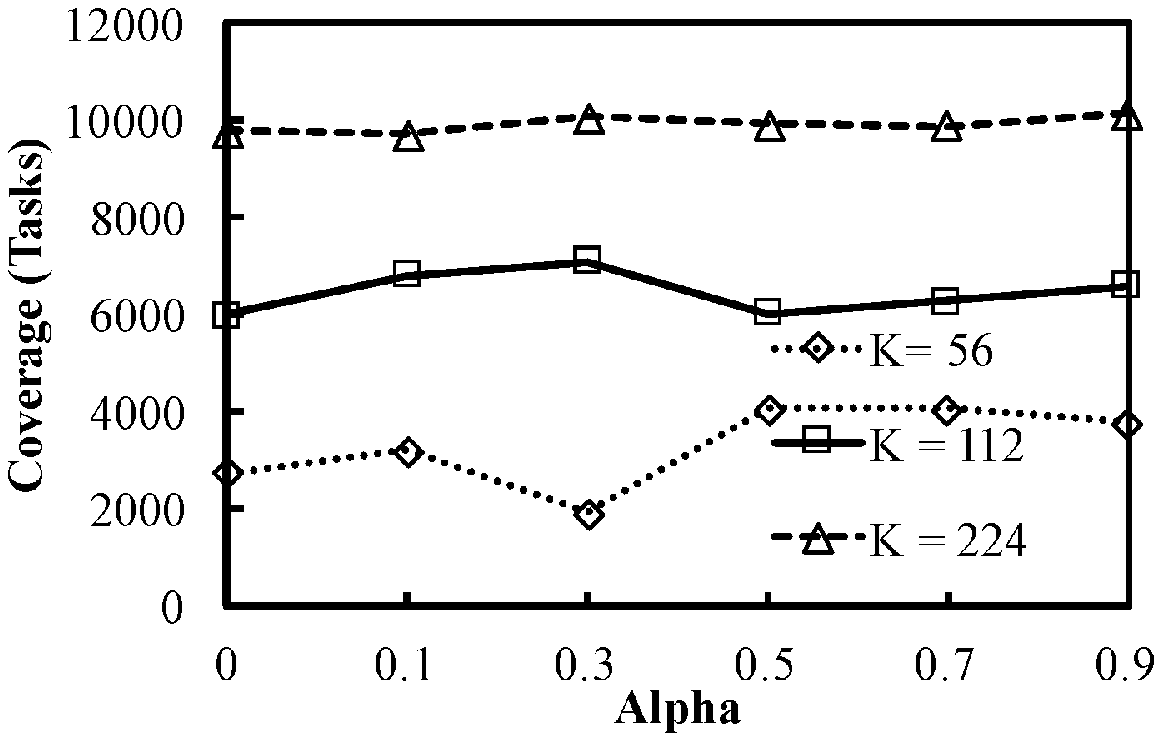}}
	\hspace{30pt}
	\subfigure[Average number of activations (vary $\alpha$)]{\label{fig:go_cosine_ol_vary_a2}\includegraphics[width=.42\textwidth]{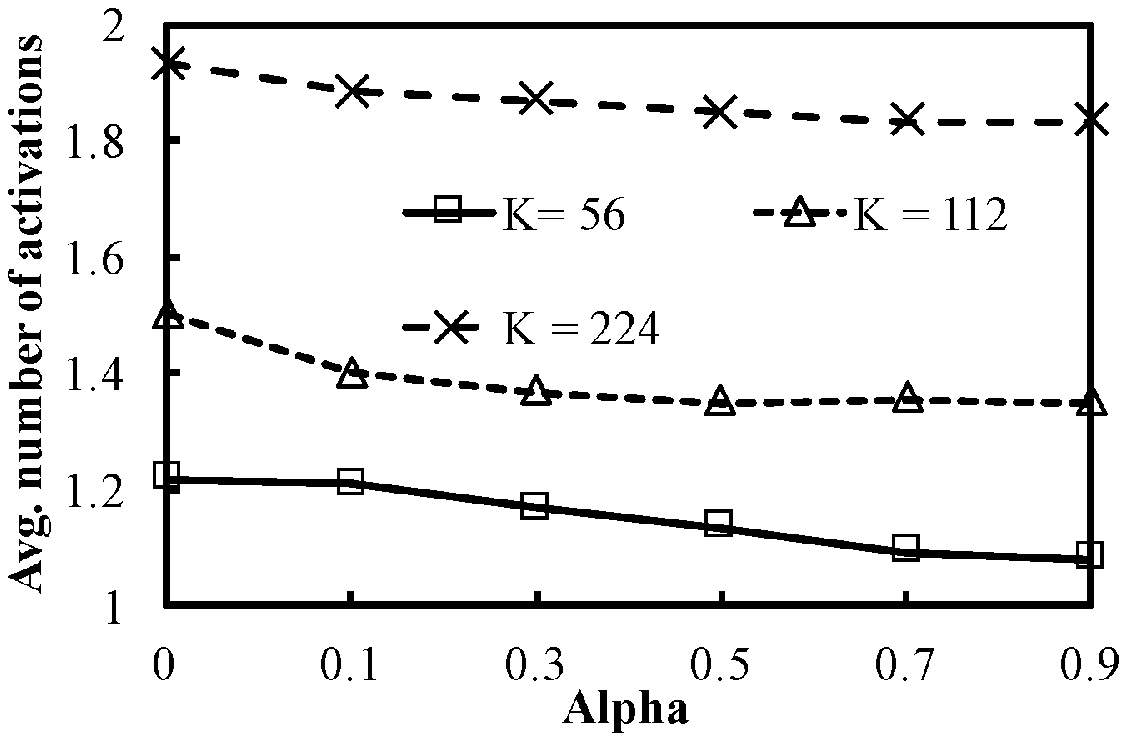}}
	\caption{Performance of \emph{EqualGA} and \emph{AdaptT-MOO} in the when varying $\alpha$.}
	\label{fig:go_overload_f1}
\end{figure}

\textbf{Worker Overload Minimization:}
In this section we evaluate the performance of the multi-objective optimization techniques in both fixed-budget and dynamic-budget scenarios. The techniques are evaluated in terms of balancing the trade-off between maximizing task coverage and minimizing worker overload.
In the fixed-budget scenario (Section~\ref{sec:ol_fixed}), \emph{EqualGA} refers to the equal-budget strategy with NSGA in Algorithm \ref{alg:ga}. We observe that varying coefficient $\alpha$ does not significantly change task coverage (Figure~\ref{fig:go_cosine_ga_vary_a}). This is due to the equal allocation of the total budget to each time period, which yields suboptimal task coverage. By increasing $\alpha$ the average number of activations per worker in Figure \ref{fig:go_cosine_ga_vary_a2}
 shows a slightly decreasing trend, due to a higher weight on the second objective.
In the dynamic-budget scenario, AdaptT-MOO refers to  adaptive budget allocation with temporal local heuristic and the multi-objective optimization in Section~\ref{sec:ol_dynamic}. Figure~\ref{fig:go_cosine_ol_vary_a} shows that the task coverage is quite stable as $\alpha$ increases while the average number of activation decreases significantly. This means that our adaptive budget allocation strategy achieves workload balancing among the workers at a very small cost  in utility. Furthermore,  without loss of generality, based on the observations, we set $\alpha = 0.1$ for the following experiments. 

\begin{figure*}[ht]
	\centering
	\subfigure[Fixed budget]{\label{fig:go_cosine_ol_ga}\includegraphics[width=.42\textwidth]{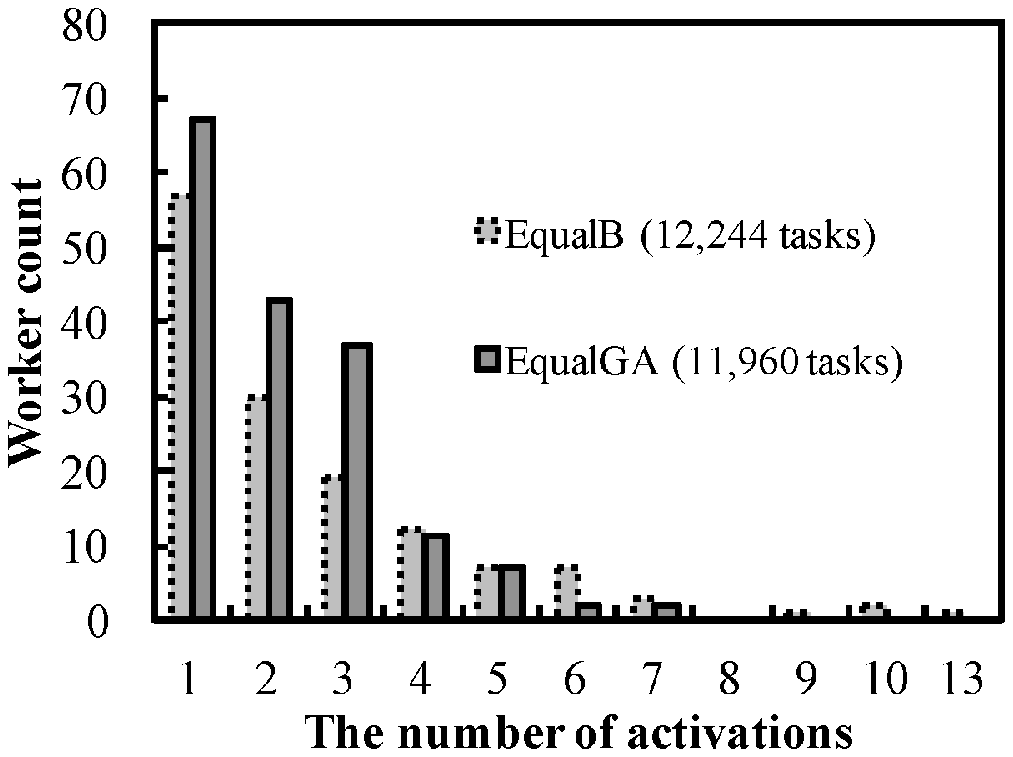}}
	\hspace{30pt}
	\subfigure[Dynamic budget]{\label{fig:go_cosine_ol_moo}\includegraphics[width=.42\textwidth]{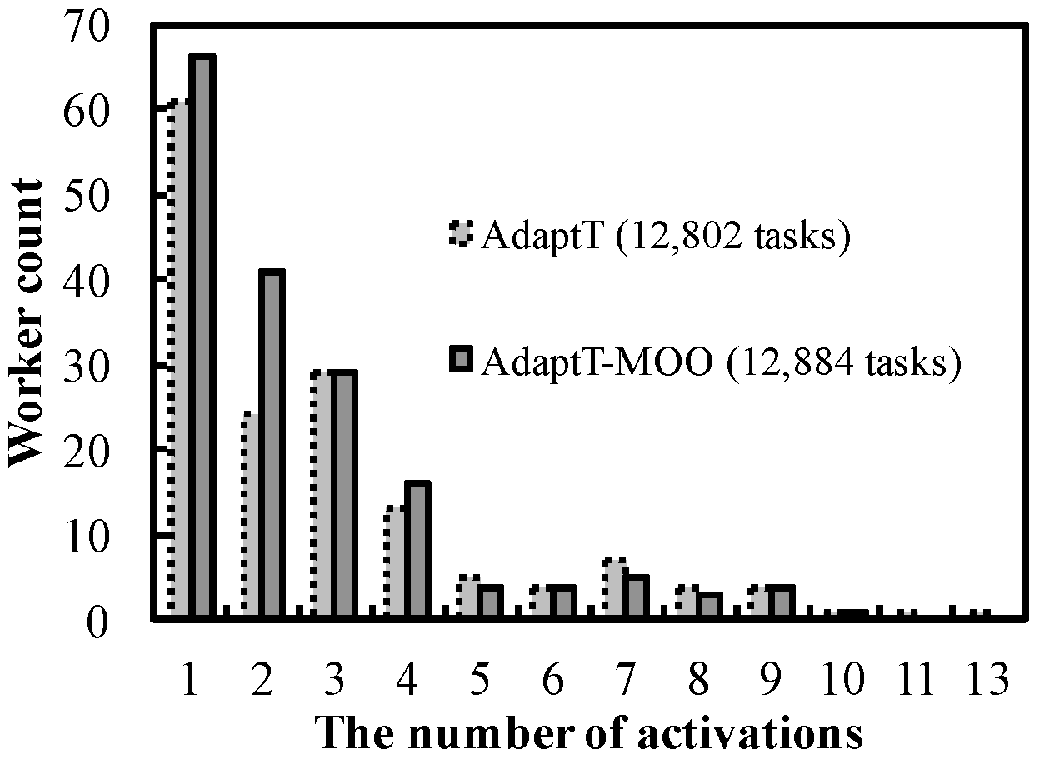}}
	\caption{Worker activation count distribution of MOO-based algorithms in the fixed and dynamic budget scenarios ($K=448, \alpha=0.1$).}
	\label{fig:go_overload_fd}
\end{figure*}

Figure \ref{fig:go_overload_fd} shows the distribution of activation counts of selected workers when $K=448$. In the fixed-budget setting, it is shown that \emph{EqualGA} does not cover as many tasks as \emph{EqualB} but it activates more workers for a small number of times, i.e., 1, 2, and 3 times. In the dynamic-budget setting, \emph{AdaptT-MOO} also has more workers with a small number of activations and yields comparable task coverage, compared to \emph{AdaptT}.  We conclude that our solutions can mitigate worker overloading without compromising task assignment. 


\subsubsection{Runtime Measurements}
\label{sec:runtim}
\begin{figure}[ht]
	\centering
		\subfigure[Fixed budget, Go-COSINE]{\label{fig:go_cosine_runtime_fixed}\includegraphics[width=.42\textwidth]{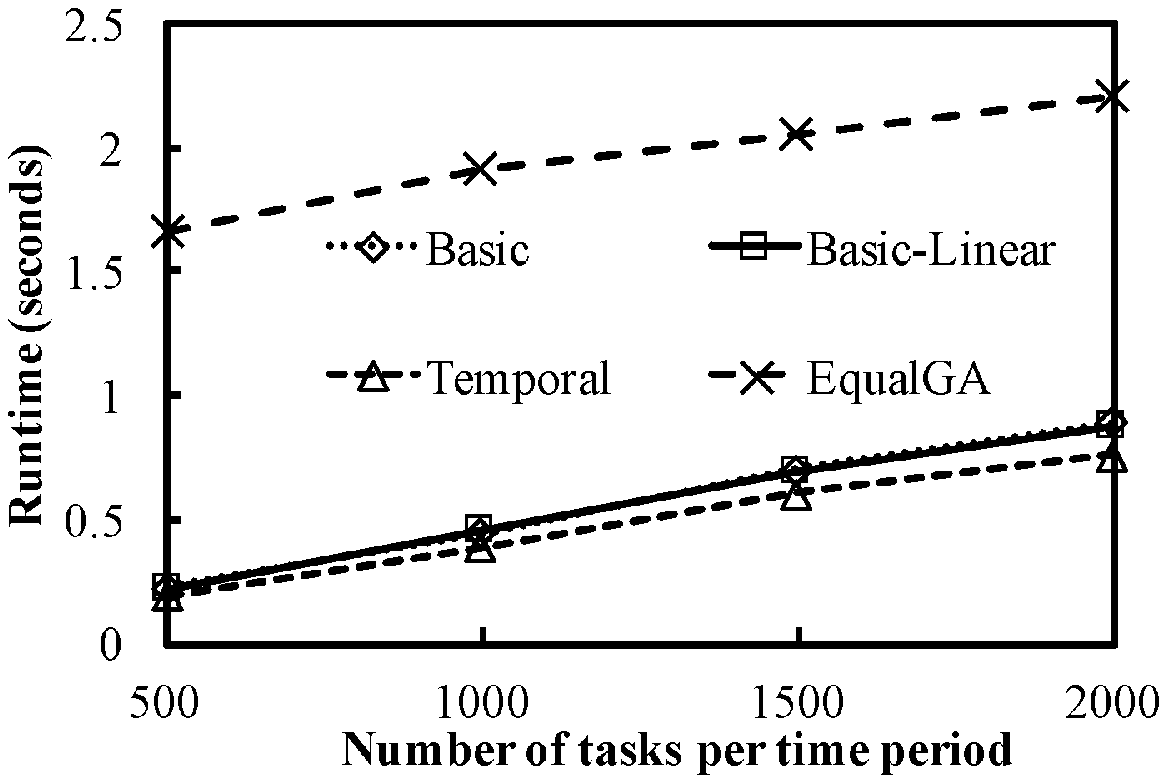}}
	\hspace{30pt}
	\subfigure[Dynamic budget, Go-COSINE]{\label{fig:go_cosine_runtime_dynamic}\includegraphics[width=.42\textwidth]{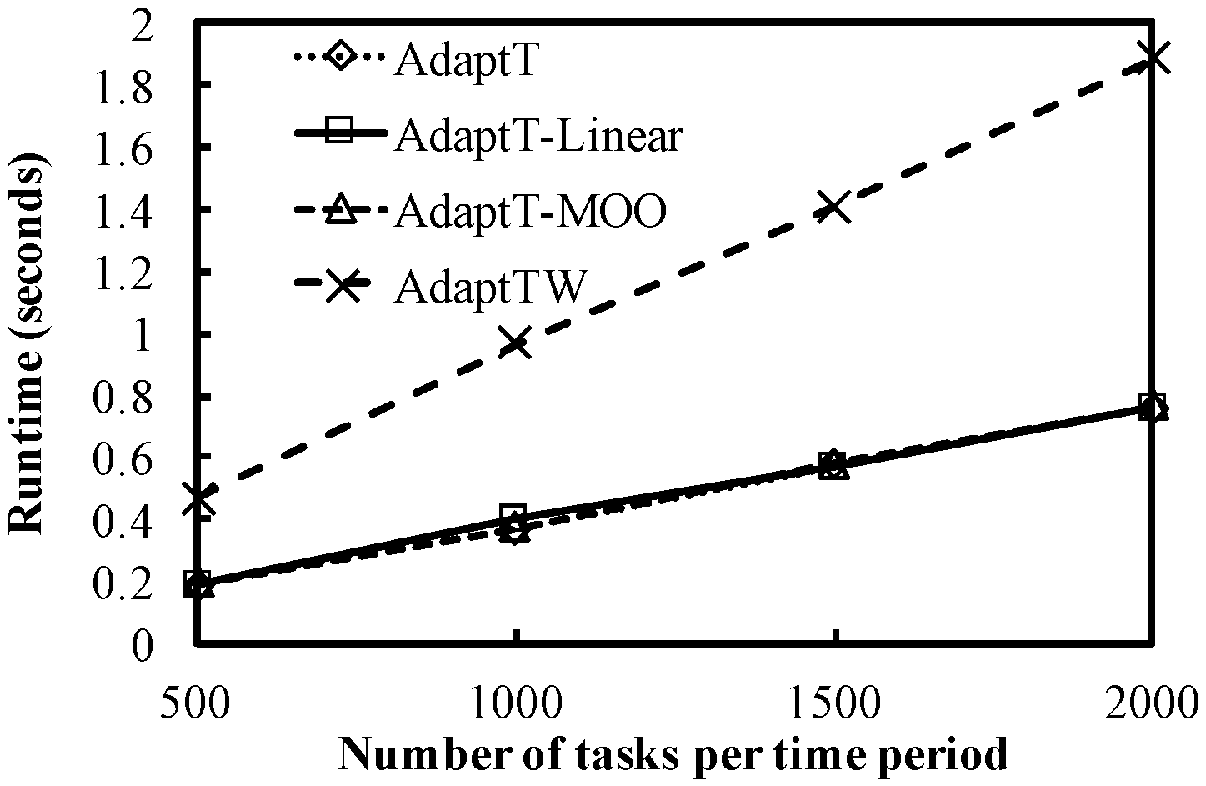}}
	
	\subfigure[Fixed budget, Fo-COSINE]{\label{fig:fo_cosine_runtime_fixed}\includegraphics[width=.42\textwidth]{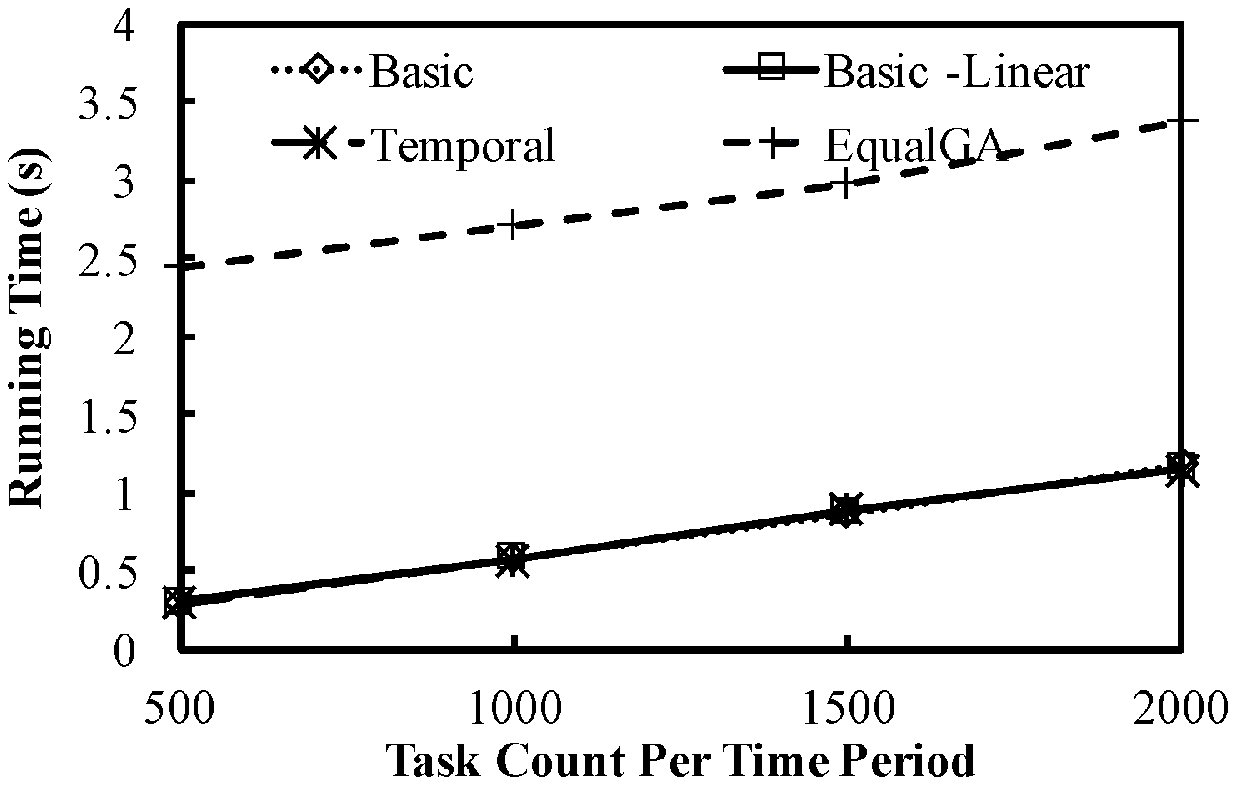}}
	\hspace{30pt}
	\subfigure[Dynamic budget, Fo-COSINE]{\label{fig:fo_cosine_runtime_dynamic}\includegraphics[width=.42\textwidth]{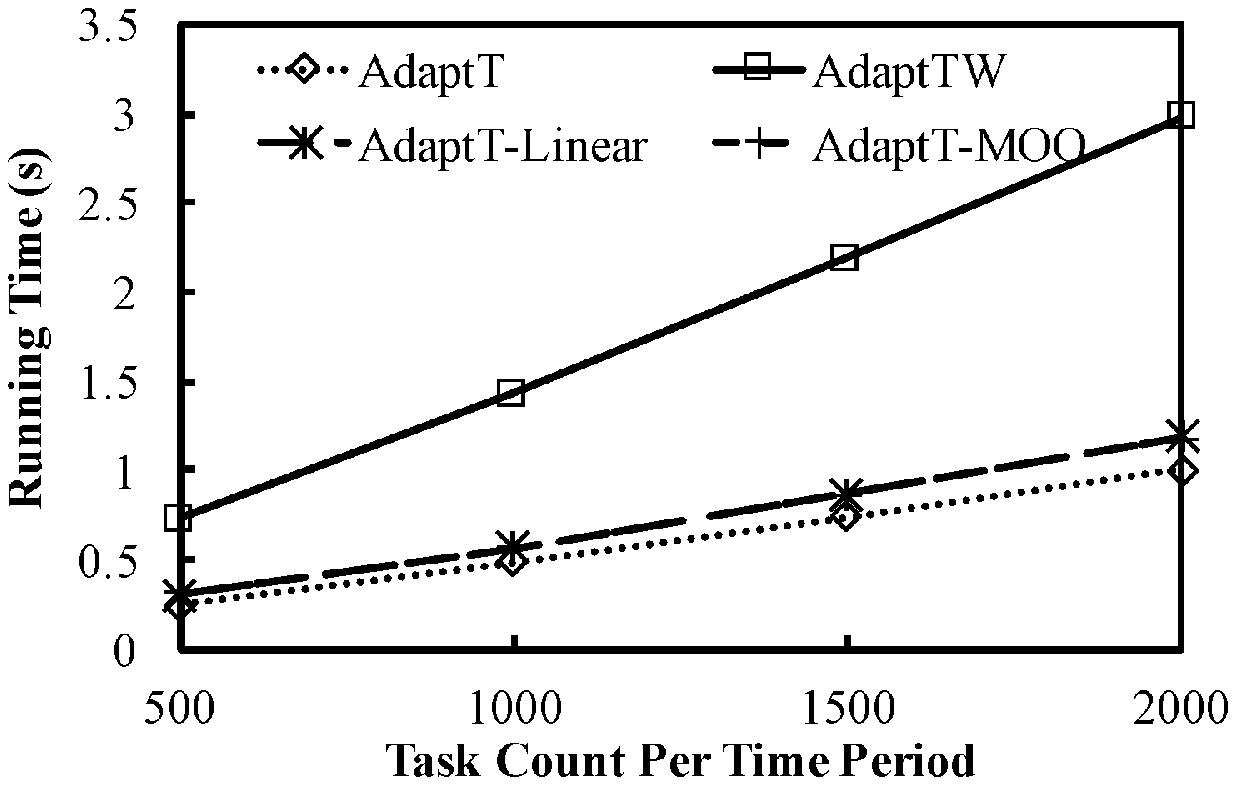}}
	\caption{Average runtime per time period with Go-COSINE and Fo-COSINE.}
	\label{fig:runtime}
\end{figure}

Figure~\ref{fig:runtime} shows the runtime performance of our online algorithms by varying the number of tasks per time period. As expected, the runtime linearly increases when the number of tasks grows. In the fixed-budget scenario,
the runtimes of the local heuristics (e.g., \emph{Temporal}) are the same as \emph{Basic} while the runtime of \emph{EqualGA} is higher due to having a large number of iterations for Algorithm~\ref{alg:ga}. We do not show the runtime of \emph{Spatial} heuristic and \emph{Zipfian} utility model but their runtimes are similar to \emph{EqualB} and \emph{EqualT-Linear}, respectively. In the dynamic-budget scenario, the runtime of \emph{AdaptTW} is higher than \emph{AdaptT, AdaptT-Linear}, and \emph{AdaptT-MOO}. This suggests that the workload heuristic significantly increases the overhead of \emph{AdaptT}. However, the MOO extension does not incur observable runtime overhead in the dynamic-budget scenario.

\section{Discussion}
\label{sec:discuss}

Existing studies show that knowing the worker mobility pattern a priori can improve the efficiency of the task assignment~\cite{ji2016urban,zhang2015event}. 
Even though, our solution does not consider individual worker mobility pattern, i.e. the worker's trajectory, for task assignment. However, our heuristics (Section~\ref{sec:spatial_heuristic}) does consider worker population mobility pattern by prioritizing tasks whose locations are not likely to be visited by many workers in the future. 
Furthermore, our dynamic budget algorithm (Algorithm~\ref{alg:adapt}) takes into account the dynamic arrivals of workers and tasks as well as their co-location relationship. 

It is worth noting that in our problem settings  1) task assignment is real-time and online and 2) workers are not required to travel to perform tasks. Workers can respond to a task immediately after receiving the task notifications from SC-server. Therefore, they do not need to perform a sequence of tasks as in a typical mobile crowdsourcing where workers often chain multiple tasks to maximize their earnings while minimizing travel time~\cite{ji2016urban}.
In addition, workers' trajectories within the same task region would not have much impact in our problem setting, as the workers are not required to travel to perform the task. Obviously, as the workers move, they may become relevant to another spatial task and/or irrelevant to the prior task, which can be represented as the addition and deletion of a worker in our framework at a given snapshot.

\section{Conclusion}
\label{sec:con}

Motivated by weather crowdsourcing applications, we introduced the problem of Hyperlocal Spatial Crowdsourcing, where tasks can be performed by workers within their spatiotemporal vicinity. We studied task assignment in Hyperlocal SC to maximize the covered tasks without exceeding the budget for activating workers.  A range of problem variants was considered, including offline vs. online, budget constraint for each time period vs. for the entire campaign, single objective vs. multiple objectives, and binary vs. distance-based utility.  We showed that the offline variants are NP-hard and proposed several local heuristics and the dynamic budget allocation for the online scenario which utilize the spatial and temporal properties of workers/tasks.  We generated spatial crowdsourcing workloads with SCAWG tool and conducted extensive experiments.   We concluded that \emph{AdaptT}, which merits the temporal local heuristic and dynamic budget allocation, is the superior technique in terms of utility and runtime.   The extensions to measure distance-based utility and to minimize worker overloading were shown to be very effective and do not impose significant runtime overhead.  
As future work, we will consider non-uniform activation cost of the workers, which represents the reputation or the compensation demand of each worker.  We will also consider assigning a task to multiple workers to improve the quality of collected data and utilizing known worker mobility patterns to boost task assignment.  





\section{Acknowledgments}
We would like to thank CHRS researchers, especially Dr. Phu Dinh Nguyen for leading the development of the iRain project: http://irain.eng.uci.edu/.

This research has been funded by NSF grants IIS-1320149, CNS-1461963, the USC Integrated Media Systems Center, and the University at Albany. Any opinions, findings, and conclusions or recommendations expressed in this material are those of the authors and do not necessarily reflect the views of any of the sponsors such as NSF. 
\vspace{-5pt}
\bibliographystyle{ACM-Reference-Format-Journals}
\bibliography{acmsmall-sample-bibfile}









\end{document}